\documentclass{lmcs}
\pdfoutput=1

\usepackage{lastpage}
\lmcsdoi{20}{3}{22}
\lmcsheading{}{\pageref{LastPage}}{}{}%
{Dec.~19,~2023}{Sep.~10,~2024}{}


\usepackage{hyperref}
\usepackage[utf8]{inputenc}
\usepackage{amsmath,amssymb,amsfonts}
\usepackage{tikz}
\usetikzlibrary{calc,intersections,decorations.pathmorphing,decorations.markings,patterns,positioning,arrows.meta,shapes}
\usepackage{graphicx}
\usepackage{bussproofs}
\usepackage{bussproofs-extra}
\usepackage{xcolor}
\usepackage{mathtools}
\usepackage[retainorgcmds]{IEEEtrantools}
\usepackage{relsize}
\usepackage[inline]{enumitem}
\usepackage{adjustbox}
\usepackage[list=off]{caption}
\usepackage{thmtools, thm-restate}
\theoremstyle{plain}\newtheorem{restate-definition}[thm]{Definition}
\theoremstyle{plain}\newtheorem{restate-lemma}[thm]{Lemma}

\newcommand{\pr}[1]{\left(#1\right)}
\newcommand{\br}[1]{\left\{#1\right\}}

\renewcommand{\restriction}{\mathord{\upharpoonright}}

\newcommand{\NP}{\text{$\mathsf{NP}$}}
\newcommand{\PSPACE}{\text{$\mathsf{PSPACE}$}}
\newcommand{\TQBF}{\text{$\mathsf{TQBF}$}}
\newcommand{\FQBF}{\text{$\mathsf{FQBF}$}}

\newcommand{\myparity}{\text{parity}}
\newcommand{\QParity}{\text{QParity}}
\newcommand{\QUParity}{\text{QUParity}}
\newcommand{\LQParity}{\text{LQParity}}
\newcommand{\MParity}{\text{MParity}}

\newcommand{\KBKF}{\text{KBKF}}
\newcommand{\KBKFlq}{\text{KBKF-lq}}
\newcommand{\KBKFlqweak}{\text{KBKF-lq-weak}}
\newcommand{\KBKFlqsplit}{\text{KBKF-lq-split}}

\newcommand{\Equality}{\text{Eq}}
\newcommand{\SquaredEquality}{\text{Eq$^2$}}
\newcommand{\SquaredEqualityWithHoles}{\text{H-Eq$^2$}}

\newcommand{\Res}{\text{Res}}
\newcommand{\MRes}{\text{M-Res}}
\newcommand{\MResW}{\text{M-ResW}}
\newcommand{\MResWE}{\text{$\MResW_\exists$}}
\newcommand{\MResWU}{\text{$\MResW_\forall$}}
\newcommand{\MResWEU}{\text{$\MResW_{\exists\forall}$}}
\newcommand{\QRes}{\text{Q-Res}}
\newcommand{\QURes}{\text{QU-Res}}
\newcommand{\LDQRes}{\text{LD-Q-Res}}

\newcommand{\LQURes}{\text{LQU-Res}}
\newcommand{\LQUplusRes}{\text{LQU$^+$-Res}}
\newcommand{\ExpRes}{\forall\text{Exp}+\Res}
\newcommand{\QCP}{\text{CP}+\forall\text{red}}
\newcommand{\IR}{\text{IR}}
\newcommand{\IRM}{\text{IRM}}

\newcommand{\eFrege}{\text{eFrege}}

\newcommand{\QBFeFrege}{\text{eFrege}+\forall\text{red}}

\newcommand{\var}{\text{var}}

\newcommand{\res}{\text{res}}

\newcommand{\leaves}{\text{leaves}}

\newcommand{\UsedConstraintInd}{\text{{\sc Uci}}}

\newcommand{\weakB}{\text{weak-B}}
\newcommand{\splitB}{\text{split-B}}

\newcommand{\xor}{\text{xor}}
\newcommand{\exor}{\text{even-parity}}

\DeclarePairedDelimiter{\card}{\lvert}{\rvert}

\tikzset{double_border/.style={draw, double, double distance=1pt,outer sep=1.2pt}}

\begin{document}

\title{QBF Merge Resolution is powerful but unnatural} 
\titlecomment{A preliminary version of this article appeared in the proceedings of the 25th International Conference on Theory and Applications of Satisfiability Testing -- SAT 2022 \cite{MS-SAT22}.}

\author[M.~Mahajan]{Meena Mahajan\lmcsorcid{0000-0002-9116-4398}}[a,b]
\author[G.~Sood]{Gaurav Sood\lmcsorcid{0000-0001-6501-6589}}[a,b]
\address{The Institute of Mathematical Sciences, Chennai, India}
\address{Homi Bhabha National Institute, Training School Complex, Anushaktinagar, Mumbai, India}
\email{meena@imsc.res.in}
\thanks{Current affiliation: G. Sood,  School of Computing and Electrical Engineering, Indian Institute of Technology Mandi, Kamand, Himachal Pradesh, India. Current Email: gauravsood@iitmandi.ac.in}

\begin{abstract}
The Merge Resolution proof system ($\MRes$) for QBFs, proposed by Beyersdorff et al.~in 2019, explicitly builds partial strategies inside refutations. The original motivation for this approach was to overcome the limitations encountered in long-distance Q-Resolution proof system ($\LDQRes$), where the syntactic side-conditions, while prohibiting all unsound resolutions, also end up prohibiting some sound resolutions. However, while the advantage of $\MRes$ over many other resolution-based QBF proof systems was already demonstrated, a comparison with $\LDQRes$ itself had remained open. In this paper, we settle this question. We show that $\MRes$ has an exponential advantage over not only $\LDQRes$, but even over $\LQUplusRes$ and $\IRM$, the most powerful among currently known resolution-based QBF proof systems. Combining this with results from Beyersdorff et al.~2020, we conclude that $\MRes$ is incomparable with $\LQURes$ and $\LQUplusRes$.

Our proof method reveals two additional and curious features about
$\MRes$: 
\begin{enumerate*}[label=(\roman*)]
	\item $\MRes$ is not closed under restrictions, and is hence not a natural proof system, and
	\item weakening axiom clauses with existential variables provably yields an exponential advantage over $\MRes$ without	weakening.
\end{enumerate*}
We further show that in the context of regular derivations, weakening axiom clauses with universal variables provably yields an exponential advantage over $\MRes$ without weakening. These results suggest that $\MRes$ is better used with weakening, though whether $\MRes$ with weakening is closed under restrictions remains open. We note that even with weakening, $\MRes$ continues to be simulated by $\QBFeFrege$ (the simulation of ordinary $\MRes$ was shown recently by Chew and Slivovsky). 
\end{abstract}

\maketitle

\section{Introduction}
Testing satisfiability of CNF formulas (the propositional SAT problem) is $\NP$-complete and is hence believed to be hard in the worst case. Despite this, modern  SAT solvers routinely solve industrial SAT instances with hundreds of thousands or even millions of variables in close to linear time \cite{Vardi-Satisfiability-CACM14,BussN-HandbookofSAT21,MLM-HandbookofSAT21}. Recently some mathematics problems, some of which were open for almost a century, have been solved by employing SAT solvers
(see \cite{HeuleK-CACM17} for a survey). This apparent disconnect between theory and practice has led to a more detailed study of the different solving techniques.

Most successful SAT solvers use a non-deterministic algorithm called conflict-driven clause learning (CDCL) \cite{MSS-CDCL-ITC99,MMZZM-CDCL-DAC01}, which is inspired by and an improvement of the DPLL algorithm \cite{DavisP-JACM60,DavisLL-CACM62}. The solvers use some heuristics to make deterministic or randomized choices for the non-deterministic steps of the CDCL algorithm. The CDCL algorithm (and the resulting solvers) can be studied by analysing a proof system called resolution. Resolution contains a single inference rule, which given clauses $x \vee A$ and $\overline{x} \vee B$, allows the derivation of clause $A \vee B$ \cite{Blake-Thesis37,Robinson-JACM60}.
To be more precise, from a run of the CDCL algorithm (or a solver) on an unsatisfiable formula, resolution refutations of the same length (as the run of the solver) can be extracted. This means that refutation size lower bounds on resolution translate to runtime lower bounds for the CDCL algorithm and the solvers based on it. See \cite{MLM-HandbookofSAT21} for more on CDCL based SAT solvers and \cite{BussN-HandbookofSAT21} for their connection to resolution.

With SAT solvers performing so well, the community has set sights on solving Quantified Boolean formulas (QBFs). Some of the variables in QBFs are quantified universally,  allowing a more succinct but also explainable encoding of many constraints. As a result, QBF solving has many more practical applications (see \cite{ShuklaBPS-ICTAI19} for a survey). However, it is $\PSPACE$-complete \cite{StockmeyerM-STOC73} and hence believed to be much harder than SAT.%

The main way of tackling QBFs in proof systems is by adapting resolution to handle universal variables. There are two major ways of doing this, which have given rise to two orthogonal families of proof systems. Reduction-based systems allow dropping a universal variable from a clause if some conditions are met --- proof systems $\QRes$ and $\QURes$ \cite{KKF-IC95,Gelder-CP12} are of this type. In contrast, expansion-based systems eliminate universal variables at the outset by expanding the universal quantifiers into conjunctions, giving a purely propositional formula --- proof systems $\ExpRes$ and $\IR$ \cite{Janota-Expansion-vs-QRes-TCS15,BCJ-ToCT-Sep2019} are of this type. It was soon observed that, under certain conditions, producing a clause containing a universal variable in both polarities (to be interpreted in a special way, not as a tautology) is not only sound but also very useful for making proofs shorter \cite{ZhangM-ICCAD02,EglyLW-LPAR13}. This led to new proof systems of both types: reduction-based systems $\LDQRes$, $\LQURes$ and $\LQUplusRes$ \cite{BJ-FMSD12,BWJ-SAT14}, and expansion-based system $\IRM$ \cite{BCJ-ToCT-Sep2019}.

Since all these proof systems degenerate to resolution on propositional formulas, lower bounds for resolution continue to hold for these systems as well. However such lower bounds do not tell us much about the relative powers and weaknesses of these systems. QBF proof complexity aims to understand this. This is done by finding formula families which have polynomial-size refutations in one system but require super-polynomial size refutations in the other system. For example, among the reduction-based and expansion-based resolution systems, $\LQUplusRes$ and $\IRM$ respectively are the most powerful and are known to be incomparable \cite{BWJ-SAT14,BCJ-ToCT-Sep2019}. 

In this paper, we study a proof system called Merge Resolution ($\MRes$). This system was proposed in \cite{BBM-JAR20} with the goal of circumventing a limitation of $\LDQRes$. The main feature of this system is that each line of the refutation contains information about partial strategies for the universal player in the standard two-player evaluation game associated with QBFs. These strategies are built up as the proof proceeds. The information about these partial strategies allows some resolution steps which are blocked in $\LDQRes$. This makes $\MRes$ very powerful --- it has short refutations for formula families requiring exponential-size refutations in $\QRes$, $\QURes$, $\ExpRes$, and $\IR$, and also in the system $\QCP$ introduced in \cite{BCMS-IC18}. However, the authors of \cite{BBM-JAR20} did not show any advantage over $\LDQRes$ --- the system that $\MRes$ was designed to improve. They only showed advantage over a restricted version of $\LDQRes$, the system reductionless $\LDQRes$. In a subsequent paper \cite{BBMPS-ToCT}, limitations of $\MRes$ were shown ---  there are formula families which have polynomial-size refutations in $\QURes$, $\LQURes$, $\LQUplusRes$ and $\QCP$, but require exponential-size refutations in $\MRes$. This, combined with the results from \cite{BBM-JAR20}, showed that $\MRes$ is incomparable with $\QURes$ and $\QCP$. More recently, it has been shown that $\QBFeFrege$ proof system p-simulates $\MRes$ \cite{ChewSlivovsky-LMCS24}. On the solving side, $\MRes$ has recently been used to build a solver, though with a different representation for strategies \cite{BlinkhornPS-SAT21}. Some variants of $\MRes$ have been studied in \cite{CS-STACS23} from a theoretical viewpoint.

In this paper, we show that $\MRes$ is indeed quite powerful, answering one of the main questions left open in \cite{BBM-JAR20}. We show that there are formula families which have polynomial-size refutations in $\MRes$ but require exponential-size refutations in $\LDQRes$. In fact, we show that there are formula families having short refutations in $\MRes$ but requiring exponential-size refutations in $\LQUplusRes$ and $\IRM$ --- the most powerful resolution-based QBF proof systems. Combining this with the results in \cite{BBMPS-ToCT}, we conclude that $\MRes$ is incomparable with $\LQURes$ and $\LQUplusRes$; see \autoref{thm:adv-over-irm} and \autoref{thm:adv-over-lquplus}. 

The power of $\MRes$ is shown using modifications of two well-known formula families: $\KBKFlq$ \cite{BWJ-SAT14} which is hard for $\MRes$ \cite{BBMPS-ToCT}, and $\QUParity$ \cite{BCJ-ToCT-Sep2019} which we believe is also hard. The main observation is that the reason making these formulas hard for $\MRes$ is the mismatch of partial strategies at some point in the refutation. This mismatch can be eliminated if the formulas are modified appropriately. The resultant formulas, called $\KBKFlqsplit$ and $\MParity$, have polynomial-size refutations in $\MRes$ but require exponential-size refutations in $\IRM$ and $\LQUplusRes$ respectively.

\begin{figure}
	\centering
	\begin{adjustbox}{max width=\textwidth}
	\begin{tikzpicture}
	
	\newcommand{\ExpResFile}{-2.5}
	\newcommand{\MergeFile}{5}
	\newcommand{\MergeFileOffset}{1.75}
	\newcommand{\LDQFile}{10.4}
	\newcommand{\QFile}{10.4}
	\newcommand{\LQUFile}{15}
	\newcommand{\QUFile}{15}
	\newcommand{\eFregeFile}{6.35}
	
	\newcommand{\MergeRank}{5.1}
	\newcommand{\MergeRankOffset}{1.5}
	\newcommand{\FregeRank}{8.275}
	\newcommand{\eFregeRank}{10}
	\newcommand{\LQUplusRank}{6.6}
	\newcommand{\LQURank}{5}
	\newcommand{\LDQRank}{1.5}
	\newcommand{\QURank}{1.7}
	\newcommand{\QRank}{0}
	\newcommand{\ExpResRank}{0.5}
	\newcommand{\IRRank}{3.55}
	\newcommand{\IRMRank}{6.6}

	\newcommand{\RecHeight}{0.8}
	\newcommand{\ExpResRecWidth}{1.4}
	\newcommand{\LQURecWidth}{1.6}
	\newcommand{\MResRecWidth}{1.8}
	
	\newcommand{\LQRecTopLeftX}{\LQUFile - \LQURecWidth}
	\newcommand{\LQRecTopLeftY}{\LQUplusRank + \RecHeight}
	\newcommand{\LQRecBotRightX}{\QUFile + \LQURecWidth}
	\newcommand{\LQRecBotRightY}{\QURank - \RecHeight}
	\newcommand{\ExpResRecTopLeftX}{\ExpResFile - \ExpResRecWidth}
	\newcommand{\ExpResRecTopLeftY}{\IRMRank + \RecHeight}
	\newcommand{\ExpResRecBotRightX}{\ExpResFile + \ExpResRecWidth}
	\newcommand{\ExpResRecBotRightY}{\ExpResRank - \RecHeight}
	\newcommand{\MRecTopLeftX}{\MergeFile - \MergeFileOffset - \MResRecWidth}
	\newcommand{\MRecTopLeftY}{\MergeRank + \MergeRankOffset + \RecHeight}
	\newcommand{\MRecBotRightX}{\MergeFile + \MergeFileOffset + \MResRecWidth}
	\newcommand{\MRecBotRightY}{\MergeRank - \MergeRankOffset - \RecHeight}
	
	\draw[fill=black!5,rounded corners = 0.15cm] (\LQRecTopLeftX, \LQUplusRank + \RecHeight) rectangle (\LQRecBotRightX, \QURank - \RecHeight);
	\draw[fill=black!5,rounded corners = 0.15cm] (\ExpResRecTopLeftX, \ExpResRecTopLeftY) rectangle (\ExpResRecBotRightX, \ExpResRecBotRightY);
	\draw[fill=black!5,rounded corners = 0.15cm] (\MRecTopLeftX, \MRecTopLeftY) rectangle (\MRecBotRightX, \MRecBotRightY);
	
	\node[draw,rounded corners] (ExpRes) at (\ExpResFile,\ExpResRank) {$\ExpRes$};
	\node[draw,rounded corners] (IR) at (\ExpResFile,\IRRank) {$\IR$};
	\node[draw,rounded corners] (IRM) at (\ExpResFile,\IRMRank) {$\IRM$};
	\node[draw,double_border, thick] (M) at (\MergeFile,\MergeRank - \MergeRankOffset) {$\MRes$};
	\node[draw,densely dashed, thick] (MWE) at (\MergeFile - \MergeFileOffset,\MergeRank) {$\MResWE$};
	\node[draw,double_border, thick] (MWU) at (\MergeFile + \MergeFileOffset,\MergeRank) {$\MResWU$};
	\node[draw,densely dashed, thick] (MWEU) at (\MergeFile,\MergeRank + \MergeRankOffset) {$\MResWEU$};
	\node[draw,rounded corners] (LDQ) at (\LDQFile,\LDQRank) {$\LDQRes$};
	\node[draw,rounded corners] (Q) at (\QFile,\QRank) {$\QRes$};
	\node[draw,rounded corners] (QU) at (\QUFile,\QURank) {$\QURes$};
	\node[draw,rounded corners] (LQU) at (\LQUFile,\LQURank) {$\LQURes$};
	\node[draw,rounded corners] (LQUplus) at (\LQUFile,\LQUplusRank) {$\LQUplusRes$};
	\node[draw,rounded corners] (QeFrege) at (\eFregeFile,\eFregeRank) {$\QBFeFrege$};
	
	\newcommand{\Thick}{0.3mm}
	\newcommand{\Thicker}{0.55mm}
	
	\draw[{Latex[scale=1.8,open]}-] (LQU) -- (LQUplus);
	\draw[-{Latex[scale=1.8,open]}] (MWEU) -- (MWE);
	
	\draw[-{Latex[open,scale=1.8]}] (MWU) -- (M);
	
	\draw[line width = \Thicker,{Latex}-] (M) -- (MWE);
	\draw[line width = \Thicker,{Latex}-] (MWU) -- (MWEU);
	\draw[{Latex[scale=1.8]}-] (ExpRes) -- (IR);
	\draw[{Latex[scale=1.8]}-] (IR) -- (IRM);
	\draw[{Latex[scale=1.8]}-] (Q) -- (LDQ);
	\draw[{Latex[scale=1.8]}-] (Q) to [out=5,in=225,looseness=1] (QU);
	\draw[{Latex[scale=1.8]}-] (LDQ) to [out=5,in=247.5,looseness=1] (LQU);
	\draw[{Latex[scale=1.8]}-] (QU) -- (LQU);
	\draw[{Latex[scale=1.8]}-] (LDQ) to [out=180,in=292.5,looseness=1.2] (IRM);
	\draw[{Latex[scale=1.8]}-] (Q) to [out=180,in=300,looseness=1] (IR);
	\node (P8) at (\ExpResFile, \ExpResRecTopLeftY - 0.1){};
	\draw[{Latex[scale=1.8]}-] (P8) to [out=37.5,in=184,looseness=1] (QeFrege);
	\node (P9) at (\LQUFile, \LQRecTopLeftY - 0.1){};
	\draw[{Latex[scale=1.8]}-] (P9) to [out=142.5,in=356,looseness=1] (QeFrege);
	\node (P10) at (\MergeFile, \MRecTopLeftY - 0.1){};
	\draw[line width = \Thicker,{Latex}-] (P10) -- (QeFrege);	
	
	\draw[line width = \Thicker,-{Latex},dashed] (MWE) -- (MWU);
	\node (P2) at (\MRecTopLeftX + 0.13, {((\MRecTopLeftY+\MRecTopLeftY+\MRecTopLeftY+\MRecBotRightY)/4}){};
	\node (P3) at (\ExpResRecBotRightX - 0.11, {((\MRecTopLeftY+\MRecTopLeftY+\MRecTopLeftY+\MRecBotRightY)/4}){};
	\draw[line width = \Thicker,-{Latex},dashed] (P2) -- (P3);
	\node (P4) at (\MRecBotRightX - 0.12, {((\MRecTopLeftY+\MRecTopLeftY+\MRecTopLeftY+\MRecBotRightY)/4}){};
	\node (P5) at (\LQRecTopLeftX + 0.125, {((\MRecTopLeftY+\MRecTopLeftY+\MRecTopLeftY+\MRecBotRightY)/4}){};
	\draw[line width = \Thicker,-{Latex},dashed] (P4) -- (P5);
	\node (P6) at (\LQRecTopLeftX + 0.12, \MergeRank - \MergeRankOffset){};
	\node (P7) at (\LQRecTopLeftX + 0.12, \MergeRank){};
	\draw[dashed,{Latex[scale=1.8]}-] (M) -- (P6);
	\draw[line width = \Thicker,dashed,{Latex}-] (MWU) -- (P7);
	
	\newcommand{\KeyVSpace}{1}
	
	\newcommand{\NodeKeyTopRank}{-1-\KeyVSpace}
	\newcommand{\NodeKeyMiddleRank}{-2-\KeyVSpace}
	\newcommand{\NodeKeyBottomRank}{-3-\KeyVSpace}
	\newcommand{\NodeKeyFile}{0.75}
	\newcommand{\NodeDescriptionOffset}{2.65}
	
	\newcommand{\KeyLeft}{6.25}
	\newcommand{\KeyOffset}{0.7}
	\newcommand{\AtoBOffset}{1.7}
	\newcommand{\DescriptionOffset}{4.25}
	
	\node[draw, align=center, rounded corners, text=white](Nat) at (\NodeKeyFile,\NodeKeyTopRank){MM};
	\node[align = left, text width = 4cm] at (\NodeKeyFile + \NodeDescriptionOffset,\NodeKeyTopRank){Natural};
	\node[draw, align=center, double_border, thick, text=white](UnNat) at (\NodeKeyFile,\NodeKeyMiddleRank){NN};
	\node[align = left, text width = 4cm] at (\NodeKeyFile + \NodeDescriptionOffset,\NodeKeyMiddleRank){Unnatural};
	\node[draw, densely dashed, thick, align=center, text=white](UnKn) at (\NodeKeyFile,\NodeKeyBottomRank){MM};
	\node[align = left, text width = 4cm] at (\NodeKeyFile + \NodeDescriptionOffset,\NodeKeyBottomRank){Unknown};
	
	\node[draw, align=center, rounded corners=0.05cm](Ai) at (\KeyLeft,\NodeKeyTopRank){A};
	\node[draw, align=center, rounded corners=0.05cm](Bi) at (\KeyLeft + \AtoBOffset,\NodeKeyTopRank){B};  
	\draw[-{Latex[scale=1.8,open]}] (Ai) -- (Bi);
	\node[align = left, text width = 4cm] at (\KeyLeft + \DescriptionOffset,\NodeKeyTopRank){A p-simulates B};
	
	\node[draw, align=center, rounded corners=0.05cm](Ai) at (\KeyLeft,\NodeKeyMiddleRank){A};
	\node[draw, align=center, rounded corners=0.05cm](Bi) at (\KeyLeft + \AtoBOffset,\NodeKeyMiddleRank){B};      
	\draw[-{Latex[scale=1.8]}] (Ai) -- (Bi);
	\node[align = left, text width = 4cm] at (\KeyLeft + \DescriptionOffset,\NodeKeyMiddleRank){A  p-simulates B; \\B does not simulate A};
	
	\node[draw, align=center, rounded corners=0.05cm](Aii) at (\KeyLeft,\NodeKeyBottomRank){A};
	\node[draw, align=center, rounded corners=0.05cm](Bii) at (\KeyLeft + \AtoBOffset,\NodeKeyBottomRank){B};
	\draw[dashed,-{Latex[scale=1.8]}](Aii)--(Bii);
	\node[align = left, text width = 4cm] at (\KeyLeft + \DescriptionOffset,\NodeKeyBottomRank){B does not simulate A};

\end{tikzpicture}
	\end{adjustbox}
	\caption{Relations among resolution-based QBF proof systems, with new results and observations highlighted using thicker lines. In addition, regular $\MResWU$ strictly p-simulates regular $\MRes$.
          Lines from a big grey box mean that the line is from every proof system within the box.
		}
	\label{fig:simulations}
\end{figure}
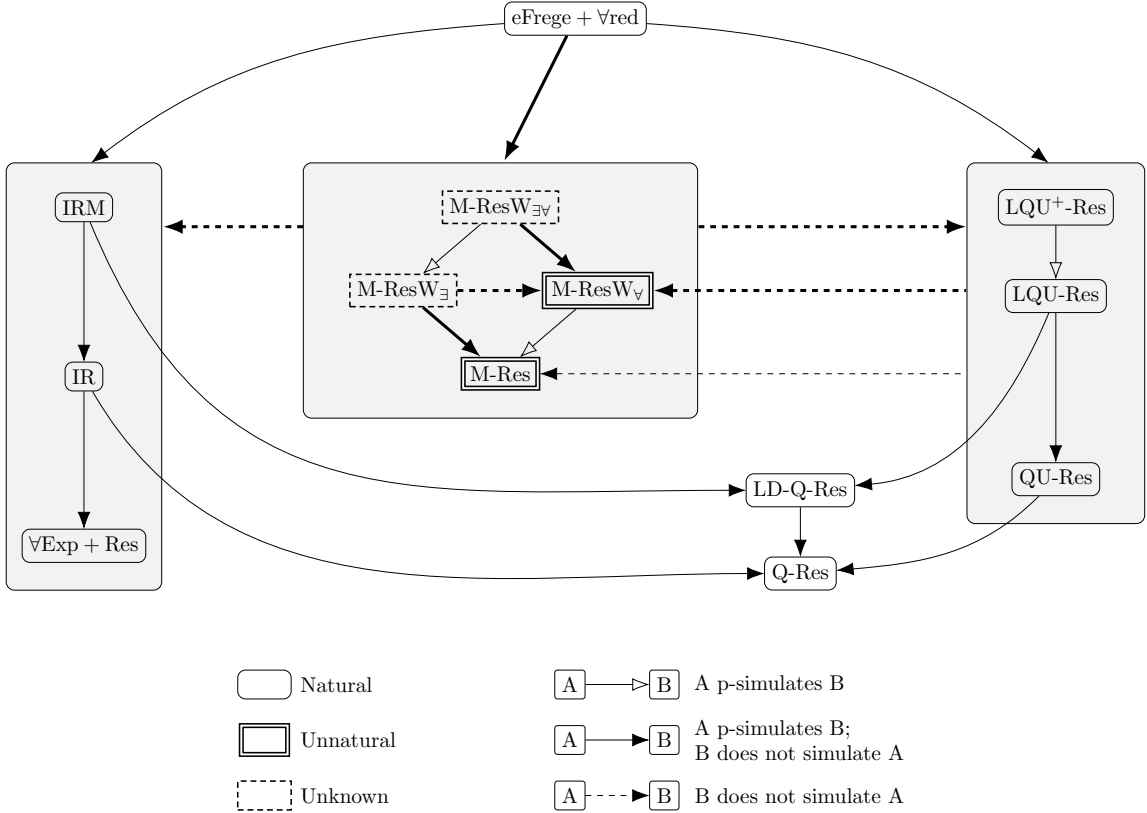

We observe that the modification of $\KBKFlq$ is actually a weakening of the clauses. This leads to an observation that weakening adds power to $\MRes$. Weakening is a rule that is sometimes augmented to resolution. This rule allows the derivation of $A \vee x$ from $A$, provided that $A$ does not contain the literal $\overline{x}$. The weakening rule is mainly used to make resolution refutations more readable --- it can not make them shorter \cite{Atserias-JACM04}. The same holds for all the known resolution-based QBF proof systems with the exception of $\MRes$ --- allowing weakening can make $\MRes$ refutations exponentially shorter. We distinguish between two types of weakenings, namely existential clause weakening and strategy weakening. Both these weakenings were defined in the original paper \cite{BBM-JAR20} in which $\MRes$ was introduced. However, these weakenings were used only for Dependency-QBFs (DQBFs); in that setting they are necessary for completeness. The potential use of weakening for QBFs was not explicitly addressed. Here, we show that existential clause weakening adds exponential power to $\MRes$; see \autoref{thm:mreswe-vs-mres}. We do not know whether strategy weakening adds power to $\MRes$. However, we show that it does add exponential power to regular $\MRes$; see \autoref{thm:strategy-weak-adds-power-to-reg-mres}. At the same time, weakening of any or both types does not make $\MRes$ unduly powerful; we show in \autoref{thm:QBFeFrege-simulates-MResW} that $\QBFeFrege$ polynomially simulates (p-simulates) $\MRes$ even with both types of weakenings added. This is proven by observing that the p-simulation of $\MRes$ in \cite{ChewSlivovsky-LMCS24} can very easily be extended to handle weakenings.

Another observation from our main result is that $\MRes$ is not closed under restrictions. Closure under restrictions is a very important property of proof systems. For a (QBF) proof system, it means that restricting a false formula by a partial assignment to some of the (existential) variables does not make the formula much harder to refute. Note that a refutation of satisfiability of a formula implicitly encodes a refutation of satisfiability of all its restrictions, and it is reasonable to expect that such refutations can be extracted without paying too large a price. This is indeed the case for virtually all known proof systems to date. Algorithmically, CDCL-based solvers work by setting some variables and simplifying the formula \cite{MLM-HandbookofSAT21}. Without closure under restrictions, setting a bad variable may make the job of refuting the formula exponentially harder.  Because of this reason, proofs systems which are closed under restrictions have been called \emph{natural proof systems} \cite{BKS-JAIR04}. We show in \autoref{thm:unnatural} that $\MRes$, with and without strategy weakening, is unnatural. We believe this would mean that it is hard to build QBF solvers based on it. On the other hand, we do not yet know whether it remains unnatural if existential clause weakening or both types of weakenings are added. We believe that this is the most important open question about $\MRes$ --- a negative answer can salvage it.

Our results are summarized in Figure \ref{fig:simulations}. 

\section{Preliminaries}
The sets $\br{1, 2, \ldots, n}$ and $\br{m, m+1, \ldots, n}$ are abbreviated as $[n]$ and $[m,n]$ respectively. A literal is a variable or its negation; a clause is a disjunction of literals.
We will interchangeably denote clauses as disjunctions of literals as well as sets of literals. A propositional formula in conjunctive normal form (cnf) is a conjunction of clauses, equivalently a set of clauses.

\subsection{Quantified Boolean Formulas}
A Quantified Boolean Formula (QBF) in prenex conjunctive normal form (p-cnf), denoted $\Phi = \mathcal{Q}. \phi$, consists of two parts:
\begin{enumerate*}[label=(\roman*)]
	\item a quantifier prefix $\mathcal{Q} = Q_1 Z_1, Q_2 Z_2, \ldots, Q_n Z_n$ where the $Z_i$ are pairwise disjoint sets of variables, each $Q_i \in \br{\exists, \forall}$, and $Q_i \neq Q_{i+1}$; and
	\item a conjunction of clauses $\phi$ with variables in $Z = Z_1 \cup \cdots \cup Z_n$.
\end{enumerate*}
In this paper, when we  say QBF, we  mean a p-cnf QBF. 
The set of existential (resp.~universal) variables of $\Phi$, denoted $X$ (resp.~$U$), is the union of $Z_i$ for which $Q_i = \exists$ (resp.~$Q_i = \forall$).

The semantics of a QBF is given by a two-player evaluation game played on the QBF. In a run of the game, the existential player and the universal player take turns setting the existential and the universal variables respectively in the order of the quantification prefix. The existential player wins the run of the game if every clause is set to true. Otherwise the universal player wins. The QBF is true (resp.~false) if and only if the existential player (resp.~universal player) has a strategy to win all potential runs, i.e.\ a winning strategy. The winning strategy for the existential (resp.~ universal) player is called a model (resp.~ countermodel).

\subsection{Proof systems}
\begin{defiC}[\cite{CR-JSL79}]
    For $L\subseteq \Sigma^*$, a proof system for $L$ is a
    polynomial-time computable function $f:\Delta^*\longrightarrow
    \Sigma^*$ whose range is exactly $L$.
    For some $x\in\Delta^*$ and $y\in\Sigma^*$, if $y=f(x)$, then $x$ is an
    $f$-proof of $y$, that is, a proof of $y$ in the proof system $f$
    (a proof that $y\in L$). The size of the proof is the length of $x$.    
\end{defiC}
The condition range$(f)=L$ is often stated in two parts:
\begin{itemize}
\item \emph{Soundness:} For any $y \in \Sigma^*$ and $x \in \Delta^*$, if $f(x)=y$, then $y \in L$.
\item \emph{Completeness:} For every $y \in L$, there exists $x \in \Delta^*$ such that $f(x)=y$.
\end{itemize}

In this paper, we will be interested in the languages of True QBFs ($\TQBF$) and False QBFs ($\FQBF$). A proof system for the language $\FQBF$ is also called a refutational system and the proofs in this system are called refutations. 

To compare the strength of different proof systems, we use the notion of simulations and $p$-simulations, Def~1.5 in \cite{CR-JSL79}; see also Def.~1.5.4 in \cite{krajicek2019}.

\begin{defi}
	Let $P$ and $Q$ be two proof systems for $QBFs$.
	\begin{itemize}
	\item
	We say that $P$ \emph{simulates} $Q$ if there is a computable function $g$ that transforms proofs in $Q$  to proofs in $P$ with at most a polynomial blow-up in size. 
	\item If, in addition, $g$ is polynomial-time computable, then we say that $P$ \emph{polynomially simulates ($p$-simulates)} $Q$.
	\item If P $p$-simulates $Q$ but $Q$ does not simulate $P$, then we say that $P$ is \emph{strictly stronger} than $Q$.
	\end{itemize} 
\end{defi}

For a formula $\Phi$ and a partial assignment $\rho$ to some of its variables,
 $\Phi\restriction_{\rho}$ denotes the restricted formula resulting from setting the specified variables according to $\rho$. 
\begin{defi}\label{def:restrictions}
	A QBF proof system $P$ is \emph{closed under restrictions} if for every false QBF $\Phi$ and every partial assignment $\rho$ to some existential variables, the size of the smallest $P$-refutation of $\Phi \restriction_{\rho}$ is at most polynomial in the size of the smallest $P$-refutation of $\Phi$.
\end{defi}

\begin{rem}
	Sometimes a stricter definition is used, requiring that a refutation of $\Phi\restriction_{\rho}$ be constructible in polynomial time from every refutation of $\Phi$. We will prove that $\MRes$ is not closed under restrictions for the weaker definition (and hence also for the stricter definition).
\end{rem}

\begin{defiC}[\cite{BKS-JAIR04}]
  \label{def:natural}
	A proof system is \emph{natural} if it is closed under restrictions.
\end{defiC}

\subsection{The Merge Resolution proof system}
Merge Resolution ($\MRes$) is a proof system for refuting false QBFs. Its original definition is rather technical and can be found in \cite{BBM-JAR20}. Here, we first give an informal description and then reproduce the definition as presented in \cite{BBMPS-ToCT}. The reader already familiar with the proof system $\MRes$ can skip this section altogether.

An $\MRes$ refutation of a QBF $\Phi = Q. \phi$ is a sequence of lines. Each line consists of an ordered pair --- the first part of the pair is a clause $C$ over the existential variables; and the second part is a set of
branching programs, $\br{M^u \mid u \in U}$, one branching program for each universal variable. These branching programs are called merge maps and represent partial strategies $h^u$ for the corresponding universal variables; the
internal nodes of the merge map $M^u$  are labelled by the existential variables to the left of $u$ in the quantifier prefix, and the leaves are labelled by $0$ and $1$.
Merge maps with no branching nodes are  called \emph{trivial merge maps}, denoted by $*$. A trivial merge map computes an `undefined' function. %

The rules of the system maintain the invariant that at every line, the set of functions $\{h^u\}$ gives a partial strategy (for the universal player) that wins whenever the existential player  plays from the set of assignments falsifying $C$. The goal is to derive the line with the empty clause; the corresponding strategy at this line will then be a countermodel.  

Each line of the refutation is either obtained from an axiom (i.e.\ a clause of $\phi$), or is obtained from two previous lines by resolution on the clauses. For axioms, the corresponding line is defined in a way that satisfies the desired invariant.  At a resolution step, if the pivot is $x$, then for universal variable $u$ right of $x$, the partial strategies are combined  via a branching on $x$. To control the size blow-up, common parts of the merge maps are identified through line numbering and are reused in the new merge map. For $u$ left of $x$, such a combination is disallowed as it would not be consistent with the semantics of the two-player evaluation game.  Thus the resolution is blocked if for any such $u$, the antecedents have different non-trivial strategies. However, if both strategies are identical, or if one of them is trivial, then carrying the non-trivial strategy  forward  maintains the invariant. Deciding whether the strategies are identical may not be easy in general, but with the chosen representation of merge maps, isomorphism is easy to check. 

We now give the formal definitions.
\begin{defi}
\emph{Merge maps} are deterministic branching programs, specified by a
sequence of instructions of one of the following two forms:
\begin{itemize}
	\item 
	$\langle$\texttt{Instruction} $\mathtt{i}\rangle$: $\mathtt{b}$, where $b \in
	\{*,0,1\}$.\footnote{In \cite{BBM-JAR20}, the notation used is
		$b \in \{*,u,\overline{u}\}$;
		$u,\overline{u},*$ denote  $u=1,u=0$, undefined respectively.
		}  \\ Merge maps containing a single such instruction are
		called simple. In particular, if $b=*$, then they are called
		trivial.
		\item
		$\langle$\texttt{Instruction} $\mathtt{i}\rangle$: \texttt{If} $\mathtt{x=0}$ \texttt{then go to} $\langle$\texttt{Instruction} $\mathtt{j}\rangle$ \texttt{else go to} $\langle$\texttt{Instruction} $\mathtt{k}\rangle$, for some $j,k <
		i$. In a merge map $M$ for $u$, all queried variables $x$ must
		precede $u$ in the quantifier prefix. \\ Merge maps with such
		instructions are called complex.
	\end{itemize}
	(All instruction numbers are positive integers.)  The merge map $M^u$ computes
	a partial strategy for the universal variable $u$ starting at the
	largest instruction number (the leading instruction) and following the
	instructions in the obvious way. The value $*$ denotes an undefined
	value.
\end{defi}
\begin{defi}
	Two merge maps $M_1$ and $M_2$ are said to be \emph{consistent}, denoted $M_1 \bowtie M_2$, if for every instruction number $i$ appearing in both
	$M_1,M_2$, the instructions with instruction number $i$ are
	identical.
\end{defi}
When two merge maps, $M_1$ and $M_2$, are consistent, it is possible to build the merge map: \texttt{If} $\mathtt{x = 0}$ \texttt{then go to} $\mathtt{M_1}$ \texttt{else go to} $\mathtt{M_2}$ without repeating the common parts of $M_1$ and $M_2$. To be more precise, the new merge map will contain all instructions of $M_1$ and $M_2$ and the following additional instruction: \texttt{If} $\mathtt{x = 0}$ \texttt{then go to} $\langle$\texttt{leading instruction of} $\mathtt{M_1}\rangle$ \texttt{else go to} $\langle$\texttt{leading instruction of} $\mathtt{M_2}\rangle$.

\begin{defi}
	Two merge maps $M_1, M_2$ are said to be \emph{isomorphic},
	denoted $M_1 \simeq M_2$, if there is a bijection between the instruction
	numbers in $M_1$ and $M_2$ transforming $M_1$ into $M_2$.
\end{defi}
Note that the isomorphism only allows renumbering the instructions, not permuting the existence variables. Also note that
isomorphic merge maps compute the same function.

\begin{defi}
	The \emph{proof system \MRes} has the following rules:
	\begin{enumerate}
		\item \emph{Axiom:} For a clause $A$ in $\phi$, let $C$ be the
		existential part of $A$. For each universal variable $u$, let
		$b_u$ be the value $u$ must take to falsify $A$; if
		$u\not\in\var(A)$, then $b_u=*$. For any natural number $i$, the
		line $(C,\{M^u:u\in U\})$ where each $M^u$ is the simple
		merge map $\langle$\texttt{Instruction} $\mathtt{i}\rangle$: $\mathtt{b_u}$ can be derived in \MRes.
		\item \emph{Resolution:} From lines $L_a=(C_a,\{M^u_a:u\in
		U\})$ for $a\in \{0,1\}$, in \MRes, the line
		$L=(C,\{M^u:u\in U\})$ can be derived, where for some
		$x\in X$,
		\begin{itemize}
			\item    $C = \res(C_0,C_1,x)$, and 
			\item for each $u\in U$: either
			(1)~$M^u_a$ is trivial and $M^u=M^u_{1-a}$ for some $a$; or
			(2)~$M^u=M^u_0\simeq M^u_1$; or
			(3)~$x$ precedes $u$, $M_1 \bowtie M_2$ and $M^u$ 
			has all the instructions of $M^u_1$ and $M^u_2$ in addition to the following leading instruction: 
			\texttt{If} $\mathtt{x = 0}$ \texttt{then go to} $\langle$\texttt{leading instruction of} $\mathtt{M^u_1}\rangle$ \texttt{else go to} $\langle$\texttt{leading instruction of} $\mathtt{M^u_2}\rangle$. 
			The instruction number of this leading instruction is the  number (position) of the line $L$ in the derivation.
		\end{itemize}
		With slight abuse of notation, we will call $L$ the resolvent of $L_0$ and $L_1$ with pivot $x$, and denote this by $L = \res(L_0,L_1,x)$. 
                
		Note that \cite{BBM-JAR20} also requires that the positive literal of the pivot  appears in the first argument, so $x \in C_0$ (i.e.~the clause at line $L_0$) and $\overline{x} \in C_1$ (the clause at line $L_1$). However, this was only for syntactic convenience, and the way we formulate our arguments, this is not necessary.)
	\end{enumerate}
\end{defi}

Note that the entire merge maps are not stored at each line, only the
leading instruction specific to the line.  Due to consistency when combining merge maps, this is
enough information to build the entire map from the derivation. As
noted in \cite{BBM-JAR20} (Proposition 19), for lines within the same
derivation, the corresponding merge maps are always
consistent. Therefore, in the above definition, we don't have to
explicitly do a consistency check.

\begin{defi}
	An $\MRes$ \emph{refutation} is a derivation using the rules of $\MRes$ and ending in a line 
	with the empty existential clause. The size of the refutation is the
	number of lines.
\end{defi}
(Note that the bit-size of representing the refutation is always polynomially bounded in the length of the refutation (the number of lines), so defining size as refutation length is acceptable.)

A refutation can be represented as a graph (with edges directed from the antecedents to the consequent, hence from the axioms to the final line). We denote the graph corresponding to refutation $\Pi$ by $G_{\Pi}$.
\begin{defi}
	Let  $Y$ be a subset of the existential variables $X$ of $\Phi$. We say that an $\MRes$
	refutation $\Pi$ of $\Phi$ is \emph{$Y$-regular} if for each $y\in Y$, there is no
	leaf-to-root path in $G_{\Pi}$ that uses $y$ as pivot more
	than once. An $X$-regular proof is simply called a \emph{regular proof}.
	If $G_\Pi$ is a tree, then we say that $\Pi$ is a \emph{tree-like proof}.
\end{defi}

For concreteness, we reproduce a simple example from \cite{BBM-JAR20}:
\begin{exa}
	For the QBF	$\exists x, \forall u, \exists t. \pr{x \vee u \vee t} \wedge \pr{\overline{x} \vee \overline{u} \vee t} \wedge \pr{x \vee u \vee \overline{t}} \wedge \pr{\overline{x} \vee \overline{u} \vee \overline{t}}$,
	here is an $\MRes$ refutation. 
	\begin{prooftree}
		\AxiomC{$x \vee t, u=0$}
		\AxiomC{$\overline{x} \vee t, u=1$}
		\BinaryInfC{$t, u=x$}
		\AxiomC{$x \vee \overline{t}, u=0$}
		\AxiomC{$\overline{x} \vee \overline{t}, u=1$}
		\BinaryInfC{$\overline{t}, u=x$}
		\BinaryInfC{$\square, u=x$}
	\end{prooftree}
	(To be pedantic, each line should contain the merge map for $u$. For simplicity, we avoid it here, describing only the function computed by the merge map.)
\end{exa}

\section{Power of Merge Resolution}
In this section, we prove that neither $\IRM$ nor $\LQUplusRes$ simulates $\MRes$, and therefore $\MRes$ has an advantage over these proof systems (as well as over $\LDQRes$, which they both simulate).

\subsection{Advantage over \texorpdfstring{$\IRM$}{IRM}}
To show that $\MRes$ is not simulated by $\IRM$, we use a variant of the well-studied $\KBKF$ formula family.  This family was first introduced in \cite{KKF-IC95}, and along with multiple variants, has been a very influential example in showing many separations. In particular, it was used to prove that $\LDQRes$ is strictly stronger than $\QRes$ \cite{EglyLW-LPAR13}. The variant 
$\KBKFlq$ was  defined in \cite{BWJ-SAT14} and used to show that $\LDQRes$ does not simulate $\QURes$. In \cite{BBMPS-ToCT}, $\KBKFlq$ was also shown to require exponentially large $\MRes$ refutations.  We reproduce the definitions of these formulas and provide some intuition about their meaning. We then define two further variants that will be useful for our purpose. 

\begin{defiC}[\cite{KKF-IC95}]\label{def:KBKF}
	$\KBKF_n$ is the QBF with the quantifier prefix \\ $\exists d_1, e_1, \forall x_1, \ldots, \exists d_n, e_n, \forall x_n, \exists f_1, \ldots, f_n$ and with the following clauses:
	\begin{align*}
		A_0 &= \br{\overline{d_1}, \overline{e_1}}\\
		A^d_i &= \br{d_i, x_i, \overline{d_{i+1}}, \overline{e_{i+1}}}& 
		A^e_i &= \br{e_i, \overline{x_i}, \overline{d_{i+1}}, \overline{e_{i+1}}} & \forall i \in [n-1]\\
		A^d_n &= \br{d_n, x_n, \overline{f_1}, \ldots, \overline{f_n}} & 
		A^e_n &= \br{e_n, \overline{x_n}, \overline{f_1}, \ldots, \overline{f_n}}\\
		B^0_i &= \br{x_i, f_i} &
		B^1_i &= \br{\overline{x_i}, f_i} & \forall i \in [n]\\
	\end{align*}
\end{defiC}

We explain below why the $\KBKF$ formulas are false. This will also provide some intuition about the meaning of the formulas. 

\begin{factC}[\cite{KKF-IC95}] \label{fact:KBKF-false}
	The $\KBKF$ formulas are false.
\end{factC}
\begin{proof}
  Consider the following strategy for the universal player: for each
  $i\in[n]$, set $x_i=d_i$. We will show that this strategy is a
  winning strategy for the universal player \footnote{Another winning
    strategy is: for all $i \in [n]$, set $x_i = \overline{e_i}$.}.
  
  To see this, we show by induction on $i$ that if the universal
  player plays according to this strategy, then the existential player
  must either set $d_i=0$ or $e_i=0$, or one of the $A_{j}$ clauses
  for some $j < i$ (either $A_j^d$ or $A_j^e$ when $j > 0$) will be
  falsified. The base case $i=1$ is immediate since $A_0$ has exactly
  the literals $\overline{d_1}, \overline{e_1}$. Consider $i\ge 2$,
  say $i=k+1$.  If some $A_j$ clause for $j < k$ is already falsified,
  then there is nothing to prove. Otherwise, by the induction
  hypothesis, at least one of $d_{k},e_{k}$ is set to $0$. Suppose
  $d_{k}=0$. Then setting $x_{k}=d_{k}=0$ reduces clause $A_k^d$ to
  $\{\overline{d_{k+1}}, \overline{e_{k+1}}\}$. Otherwise, $d_k=1$, and
  by induction, $e_k=0$. Now, setting $x_k=d_k=1$ reduces clause
  $A_k^e$ to $\{\overline{d_{k+1}}, \overline{e_{k+1}}\}$. Either way,
  the existential player must set one of $\overline{d_{k+1}},
  \overline{e_{k+1}}$ to 0, or falsify an $A_k$ clause.

  If the existential player has not yet lost the game after $x_n$ is
  set, then we know that either $d_n$ or $e_n$ is set to $0$. To
  satisfy $A_n^d$ and $A_n^e$, the existential player must set $f_i=0$
  for some $i\in [n]$. This falsifies one of $B_i^0, B_i^1$, and the
  existential player loses the game.
\end{proof}

The $\KBKFlq$ formulas are obtained from $\KBKF$ by adding some negated $f$ literals to some clauses. This is done to make the formulas hard for the $\LDQRes$ proof system by blocking resolution steps that would otherwise be allowed.
\begin{restatable}[\cite{BWJ-SAT14}]{defiC}{KBKFlqDefinition} \label{def:KBKFlq}
	$\KBKFlq_n$ is the QBF with the quantifier prefix \\ $\exists d_1, e_1, \forall x_1, \ldots, \exists d_n, e_n, \forall x_n, \exists f_1, \ldots, f_n$ and with the following clauses:
	\begin{align*}
		A_0 &= \br{\overline{d_1}, \overline{e_1}, \overline{f_1}, \ldots, \overline{f_n}}\\
		A^d_i &= \br{d_i, x_i, \overline{d_{i+1}}, \overline{e_{i+1}}, \overline{f_1}, \ldots, \overline{f_n}}& 
		A^e_i &= \br{e_i, \overline{x_i}, \overline{d_{i+1}}, \overline{e_{i+1}}, \overline{f_1}, \ldots, \overline{f_n}} & \forall i \in [n-1]\\
		A^d_n &= \br{d_n, x_n, \overline{f_1}, \ldots, \overline{f_n}} & 
		A^e_n &= \br{e_n, \overline{x_n}, \overline{f_1}, \ldots, \overline{f_n}}\\
		B^0_i &= \br{x_i, f_i, \overline{f_{i+1}}, \ldots \overline{f_n} } &
		B^1_i &= \br{\overline{x_i}, f_i, \overline{f_{i+1}}, \ldots \overline{f_n} } & \forall i \in [n-1]\\
		B^0_n &= \br{x_n, f_n} & 
		B^1_n &= \br{\overline{x_n}, f_n}
	\end{align*}
\end{restatable}

The proof of \autoref{fact:KBKF-false} can easily be modified to show that the $\KBKFlq$ formulas are false. (Consider any run of the game, where all variables are set, and the universal player has set each $x_i$ to $d_i$. If all $B$ clauses are satisfied, then working backwards we see that each $f_i$ must have been set to 1. So the $f$ literals cannot satisfy any of the $A$ clauses. Now, working backwards again through the $A$ clauses, we see that if some $d_i$ and $e_i$ are not both  set to 1, then an $A_i$ clause is falsified, and otherwise $A_0$ is falsified.)

We now define two new formula families: $\KBKFlqweak$ and $\KBKFlqsplit$. 

\begin{defi}\label{def:KBKFlqweak}
	$\KBKFlqweak$ has the same quantifier prefix as $\KBKF$, and all the $A$-clauses of $\KBKFlq$, but it has the following clauses instead of $B^0_i$ and $B^1_i$:
	\begin{equation*}
		\begin{rcases}
			\weakB^0_i &= d_i \vee B^0_i \;\\
			\weakB^1_i &= \overline{d_i} \vee B^1_i
		\end{rcases}
		\qquad \forall i \in [n]
	\end{equation*}
\end{defi}

\begin{defi}\label{def:KBKFlqsplit}
	$\KBKFlqsplit$ has all variables of $\KBKFlq$ and one new variable $t$ quantified existentially in the first block, so its quantifier prefix is $\exists t, \exists d_1, e_1, \forall x_1, \ldots, \exists d_n, e_n, \forall x_n,\allowbreak \exists f_1, \ldots, f_n$. It has all the $A$-clauses of $\KBKFlq$, but the following clauses instead of $B^0_i$ and $B^1_i$:
	\begin{equation*}
		\begin{rcases}
			\splitB^0_i &= t \vee B^0_i \;\\
			\splitB^1_i &= t \vee B^1_i \\
			T^0_i &= \br{\overline{t}, d_i} \\
			T^1_i &= \br{\overline{t}, \overline{d_i}} \\
		\end{rcases} 
		\qquad \forall i \in [n]
	\end{equation*}
\end{defi}
It is straightforward to see that both these formulas are false as well.
With the universal player's strategy of setting $x_i=d_i$, each 
$\weakB^b_i$ clause is effectively $B^b_i$, so the strategy is a winning strategy for $\KBKFlqweak$ as well. In $\KBKFlqsplit$, if $t$ is set to 1, then one clause in each  $T_i^0,T_i^1$ pair is falsified. Otherwise, the remaining formula is the same as $\KBKFlq$, which we know is false.

\begin{lem} \label{lem:kbkfweak-mres-poly}
	$\KBKFlqweak$ has polynomial-size $\MRes$ refutations.
\end{lem}
\begin{proof}
We use the clause-names $A_0, A^d_i, \weakB^0_i$ etc.\ to denote the clause, merge map pair corresponding to the respective axioms.
  
Let $L''_i$ denote the $\MRes$-resolvent of $\weakB^0_i$ and
$\weakB^1_i$. It has only one non-trivial merge map, setting
$x_i=d_i$.  Starting with $A_0$, resolve in sequence with $A_1^e$,
$A_1^d$, $A_2^e$, $A_2^d$, and so on up to $A_n^e$, $A_n^d$ to derive
the line with all negated $f$ literals and merge maps computing
$x_i=d_i$ for each $i$.  Now sequentially resolve this with $L''_1$,
$L''_2$, up to $L''_n$ to obtain the empty clause. It can be verified that
none of these resolutions are blocked, and the final merge maps compute
the winning strategy $x_i=d_i$ for each $i$.

The refutation is pictorially depicted  below. 
{\footnotesize 
	\begin{prooftree}
		\AxiomC{$A_0$}
		\AxiomC{$A_1^e$}
		\BinaryInfC{$L_1^e$}
		\AxiomC{$A_1^d$}
		\BinaryInfC{$L_1^d$}
		\ddotsDeduce
		\DeduceC{$\fCenter L_{n-1}^d$}
		\AxiomC{$A_n^e$}
		\BinaryInfC{$L_{n}^e$}
		\AxiomC{$A_n^d$}
		\BinaryInfC{$L'_1$}
		\AxiomC{$\weakB^0_1$}
		\AxiomC{$\weakB^1_1$}
		\BinaryInfC{$L''_1$}
		\BinaryInfC{$L'_2$}
		\ddotsDeduce
		\DeduceC{$\fCenter L'_{n}$}
		\AxiomC{$\weakB^0_n$}
		\AxiomC{$\weakB^0_n$}
		\BinaryInfC{$L''_{n}$}
		\BinaryInfC{$\pr{\Box, \br{x_1 = d_1, \ldots, x_n = d_n}}$}
	\end{prooftree}}
Here $L_i^e, L_i^d, L'_i, L''_i$, for all $i \in [n]$, are the following lines, with only non-trivial merge maps written explicitly:
\begin{itemize}
	\item $L_1^e = \pr{\br{\overline{d_1}, \overline{d_{2}}, \overline{e_{2}}, \overline{f_1}, \ldots, \overline{f_n}}, \br{x_1 = 1}}$
	\item $L_i^e = \pr{\br{\overline{d_i}, \overline{d_{i+1}}, \overline{e_{i+1}}, \overline{f_1}, \ldots, \overline{f_n}}, \br{x_1 = d_1, \ldots, x_{i-1} = d_{i-1},  x_i = 1}}$ for all $i \in [2,n-1]$
	\item $L_n^e = \pr{\br{\overline{d_n}, \overline{f_1}, \ldots, \overline{f_n}}, \br{x_1 = d_1, \ldots, x_{n-1} = d_{n-1},  x_n = 1}}$
	\item $L_i^d = \pr{\br{\overline{d_{i+1}}, \overline{e_{i+1}}, \overline{f_1}, \ldots, \overline{f_n}}, \br{x_1 = d_1, \ldots, x_i = d_i}}$ for all $i \in [n-1]$
	\item $L'_i = \pr{\br{\overline{f_{i}}, \ldots, \overline{f_n}}, \br{x_1 = d_1, \ldots, x_n = d_n}}$ for all $i \in [n]$
	\item $L''_i =  \pr{\br{f_i, \overline{f_{i+1}}, \ldots \overline{f_n}}, \br{x_i = d_i}}$ for all $i \in [n-1]$
	\item $L''_n =  \pr{\br{f_n}, \br{x_n = d_n}}$ \qedhere
\end{itemize}
\end{proof}

Observe that the extra $d_i$ variable in $\weakB^0_i$ and $\weakB^1_i$ (in contrast to $B^0_i$ and $B^1_i$ in $\KBKFlq$) allows us to resolve these two lines. This gives the clause $L_i''$ %
whose merge map computes $x_i=d_i$. This merge map is isomorphic to the merge map for $x_i$ in the line derived by resolving the $A_i$ lines. This isomorphism allows the polynomial-size refutation.

\begin{lem}\label{lem:kbkfsplit-mres-poly}
	$\KBKFlqsplit$ has polynomial-size $\MRes$ refutations.
\end{lem}
\begin{proof}
  For each $i\in[n]$ and $k \in \br{0,1}$, resolving $\splitB^k_i$ and $T^k_i$ yields   $\weakB^k_i$. This gives us the $\KBKFlqweak$ formula family which, as shown in \autoref{lem:kbkfweak-mres-poly}, has polynomial-size $\MRes$ refutations.
\end{proof}

\begin{thm}\label{thm:adv-over-irm}
  $\IRM$ does not simulate $\MRes$.
\end{thm}
\begin{proof}
The $\KBKFlqsplit$ formula family witnesses the separation. 
By \autoref{lem:kbkfsplit-mres-poly}, it has  polynomial-size $\MRes$ refutations. 
Restricting it by setting $t=0$ gives the family $\KBKFlq$, which requires exponential size to refute in $\IRM$, \cite{BCJ-ToCT-Sep2019}. Since  $\IRM$ is closed under restrictions (Lemma 11 in \cite{BCJ-ToCT-Sep2019}), $\KBKFlqsplit$  also requires exponential size to refute in $\IRM$.	
\end{proof}

\subsection{Advantage over \texorpdfstring{$\LQUplusRes$}{LQU+Res}}
To show that $\LQUplusRes$ does not simulate $\MRes$, we need a formula family which has polynomial-size refutations in $\MRes$ but requires exponential-size refutations in $\LQUplusRes$. We define a new formula family called $\MParity$, as a modification of the $\QParity$ formula family \cite{BCJ-ToCT-Sep2019}. The polynomial-size $\MRes$ refutation of $\MParity$ is obtained by mimicking the $\LDQRes$ refutation of $\QParity$ with some modifications. We then show that $\MParity$ requires exponential-size $\LQUplusRes$ refutations. Since $\LQUplusRes$ polynomially simulates $\LDQRes$ and $\LQURes$, we get the non-simulation result with respect to these proof systems also.

Let us first give a brief history of $\QParity$ and other formulas based on it. $\QParity$ was first defined in \cite{BCJ-ToCT-Sep2019} and was used to show that $\QRes$ does not p-simulate $\ExpRes$ \cite{BCJ-ToCT-Sep2019} and $\LDQRes$ \cite{LeroyChew-Thesis}. (A subsequent elegant argument in \cite{BBH-LMCS19} reproved its hardness for $\QURes$ and $\QCP$.)
The  variant  $\LQParity$, also defined in \cite{BCJ-ToCT-Sep2019}, was used to show that $\LDQRes$ does not p-simulate $\ExpRes$. Finally, the variant $\QUParity$, built by duplicating the universal variable of $\LQParity$, was used to show that $\LQUplusRes$ does not p-simulate $\ExpRes$.

We give the definition of $\QParity$, informally describe the variants $\LQParity$ and $\QUParity$, and then define our new variant $\MParity$.
We will use the following notation. For variables $o,o_1,o_2$, let $\exor(o_1,o)$ and $\exor(o_1,o_2,o)$ be the following sets of clauses\footnote{In some prior papers considering this formula, these clause sets are denoted as $\xor$. However in the wider circuit/Boolean-function-complexity community, $\xor$ or $\myparity$ refer to odd parity. For clarity, we make the condition explicit by saying $\exor$.}:
\begin{IEEEeqnarray*}{rCl}
	\exor(o_1,o) &=& \{\overline{o_1} \vee o, o_1 \vee \overline{o} \},\\
	\exor(o_1,o_2,o) &=& \{\overline{o_1} \vee \overline{o_2} \vee \overline{o}, \overline{o_1} \vee o_2 \vee o, o_1 \vee \overline{o_2} \vee o, o_1 \vee o_2 \vee \overline{o} \}
\end{IEEEeqnarray*}
We note that $\exor$ of a list of variables is just the CNF representation of the constraint that the number of variables set to `True' is even. That is, $\exor(o_1,o)$ is satisfied iff $o \equiv o_1 \pmod{2}$, and $\exor(o_1,o_2,o)$ is satisfied iff $o \equiv o_1 + o_2 \pmod{2}$.
\begin{defiC}[\cite{BCJ-ToCT-Sep2019}]
	$\QParity_n$ is the QBF $\exists x_1, \ldots , x_n, \forall z, \exists t_1, \ldots , t_n.\ \pr{\wedge_{i \in [n+1]} \zeta_{i}}$
	where:
	\begin{IEEEeqnarray*}{cCl}
		\zeta_{1} &=& \exor(x_1,t_1);\\
		\zeta_{i} &=& \exor(t_{i-1},x_i,t_i), \qquad \forall i \in [2,n];\\
		\zeta_{n+1} &=& \{t_n \vee z, \overline{t_n} \vee \overline{z}\}.
	\end{IEEEeqnarray*}
\end{defiC}

With the same quantifier prefix, replacing each clause $C$ of $\QParity$ that does  not contain $z$ with the two clauses $C\vee z$ and $C \vee \overline{z}$ gives the family $\LQParity$.

To obtain $\QUParity$, the universal variable is duplicated. That is,
the block $\forall z$ is replaced with the block $\forall z_1,
z_2$. Each clause of the form $C \vee z$ in $\LQParity$ is replaced
with the clause $C \vee z_1 \vee z_2$, and each clause of the form
$C \vee \overline{z}$ is replaced with the clause $C \vee \overline{z_1} \vee \overline{z_2}$.

It is easy to see why these formulas are false: in $\QParity$, satisfying the $\zeta_i$ clauses for $i\in [n]$ forces $t_n$ to take exactly the value $\sum_ix_i \mod 2$. Since $z$ is universally quantified, setting $z$ to the opposite value
will falsify one of the $\zeta_{n+1}$ clauses. The
tweaks to obtain $\QUParity$ and $\LQParity$ do not alter this; the same strategy (duplicated for $z_1,z_2$) remains a winning strategy for the universal player.

The short $\LDQRes$ refutation of $\QParity$ (from \cite[p.~54]{LeroyChew-Thesis}) relies on the fact that most axioms do not have universal variable $z$. This enables steps in which a merged literal $z^*$ is present in one antecedent but there is no literal over $z$ in the other antecedent. $\LQParity$ is created from $\QParity$ by replacing each clause $C$ not containing $z$ by two clauses $C \vee z$ and $C \vee \overline{z}$. Since every axiom of $\LQParity$ (and hence also each derived clause) now has a literal over $z$, we can no longer resolve clauses containing the merged literal $z^*$ with any other clause. This forbids the creation of merged literals, which in turn, forbids all possible short refutations. The same problem seems to occur in $\MRes$ also. In an $\MRes$ refutation, we have merge maps instead of (starred and unstarred) universal literals, and resolution steps are allowed if the merge maps at the antecedents are isomorphic. However, we do not know of any way of converting the constant merge maps at the axioms to merge maps which pass the isomorphism checks in later steps of the refutation. We solve this problem by defining a new formula family called $\MParity$. The $\MParity$ family is obtained from $\QUParity$ by modifying some clauses and adding some auxiliary clauses.  The auxiliary clauses help in converting the constant merge maps at the axioms of the original clauses to merge maps that pass the isomorphism tests.

\begin{defi}\label{def:mparity}
	$\MParity_n$ is the following QBF:
	\begin{equation*}
	  \mathop{\exists}_{i,j \in [n]} a_{i,j}, \exists x_1, \ldots , x_n, \forall z_1, z_2, \exists t_1, \ldots , t_n.\
          \left(
          \underbrace{\pr{\bigwedge_{i \in [n+1]} \psi_{i}}}_\psi
          \wedge 
          \underbrace{\pr{\bigwedge_{i \in [n-1]} \delta_{i}}}_\delta
          \right)
	\end{equation*}
        where $\psi$ and $\delta$ are defined as follows:
	\begin{itemize}
		\item for all $C \in \exor \pr{x_1,t_1}$, $\psi_1$ consists of the following clauses: \\ $A^0_{1,C} = C \cup \br{z_1, z_2, a_{1,n} }$ and $A^1_{1,C} = C \cup \br{\overline{z_1}, \overline{z_2}, a_{1,n} }$,
		\item for all $i \in [2,n-1]$, for all $C \in \exor \pr{t_{i-1},x_i,t_i}$, $\psi_i$ consists of the following clauses: \\ $A^0_{i,C} = C \cup \br{z_1, z_2, a_{i,n}}$ and $A^1_{i,C} = C \cup \br{\overline{z_1}, \overline{z_2}, a_{i,n}}$,
		\item for all $C \in \exor \pr{t_{n-1},x_n,t_n}$, $\psi_n$ consists of the following clauses: \\ $A^0_{i,C} = C \cup \br{z_1, z_2}$ and $A^1_{i,C} = C \cup \br{\overline{z_1}, \overline{z_2}}$, 
		\item $\psi_{n+1}$ consists of the clauses $\br{t_n, z_1, z_2}$ and 	$\br{\overline{t_n}, \overline{z_1}, \overline{z_2}}$, and
               \item 
             for all $i \in [n-1]$, $\delta_i$ consists of the following clauses:
		\begin{align*}%
			B^0_{i,j} &= \br{\overline{a_{i,j}}, x_{j}, a_{i,j-1}},& 
			B^1_{i,j} &= \br{\overline{a_{i,j}}, \overline{x_{j}}, a_{i,j-1}} \qquad \forall j \in \br{n,n-1, \ldots,i+2}\\
			B^0_{i,i+1} &= \br{\overline{a_{i,i+1}}, x_{i+1}},& 
			B^1_{i,i+1} &= \br{\overline{a_{i,i+1}}, \overline{x_{i+1}}}
		\end{align*}
        \end{itemize}
\end{defi}

To see why these formulas are false, note that satisfying all the
$\delta$ clauses requires all $a_{i,j}$, $1 \le i < j \le n$ to be set
to $0$. (Consider resolution on the $x$ variables in these clauses.) At this setting, all the  $\psi$ clauses give back $\LQParity$.

We can adapt the $\LDQRes$ refutation of $\QParity$ to an $\MRes$ refutation of $\MParity$.  We describe below exactly how this is achieved. 
The family $\MParity$ consists of two sets of clauses: $\psi$ and $\delta$. The proof has two stages. In the first stage, the $a_{i,j}$ variables are eliminated from the clauses in $\psi$ using the clauses in $\delta$. The role of these $a_{i,j}$ variables and the clauses of $\delta$ is to build up complex merge maps meeting the isomorphism condition, so that subsequent resolution steps are enabled. In the second phase, the $\LDQRes$ refutation of $\QParity$ is mimicked, eliminating the $t$ variables.

(In the proofs below, notice that each line contains a single merge map. This is done because the merge maps for $z_1$ and $z_2$ in every line are same. So, we write them only once to save space.)

For $i \in [n+1]$, let $g_i$ be the parity function $\oplus_{j\ge i}x_j$, 
and let $h_i$ denote its complement; thus $h_i$ is $\exor$ on the variables $x_i,\ldots,x_n$. 
(The parity of an empty set of variables is $0$; thus $g_{n+1}=0$ and $h_{n+1}=1$.) 
Let $M^1_i$ (resp.~$M^0_i$) be the smallest merge map which queries variables in the order $x_{i}, \ldots, x_{n}$ and computes the function $g_i$
(resp.~$h_i$). Note that both these merge maps have $2(n-i)+1$ internal nodes and two leaf nodes labelled $0$ and $1$.

The main idea is to replace the constant merge maps in the axioms of $A^0_{i,C}$ and $A^1_{i,C}$ by the merge maps $M_{i+1}^0$ and $M_{i+1}^1$ --- the set of clause, merge map pairs so generated will be denoted by $\widetilde{\psi_i}$ (and are defined below). These merge maps will allow us to pass the isomorphism checks later in the proofs. 

For $i \in [n]$, let $\widetilde{\psi_i}$ be the following sets of clause, merge map pairs:
\begin{align*}
	\widetilde{\psi_{i}~} &= \br{\pr{C,M^b_{i+1}} \mid C \in \exor(t_{i-1},x_i,t_i), b \in \{0,1\}} \qquad \forall i \in [2,n]\\
	\widetilde{\psi_{1}} &= \br{\pr{C,M^b_2} \mid C \in \exor(x_1,t_1), b \in \{0,1\}}
\end{align*}

\begin{lem} \label{lem:mparity-add-complex-strategies}
	Let the quantifier prefix be as in the definition of $\MParity$.
	Then, for all $i \in [n]$, 
	 $\psi_{i} \wedge \delta_i \vdash_{\MRes} \widetilde{\psi_{i}}$. Moreover the size of these derivations is polynomial in $n$.
\end{lem}
\begin{proof}
  At $i=n$, $\widetilde{\psi_n}$ is the same as $\psi_n$ so there is nothing to prove.

Consider now an $i\in[n-1]$.
For each $b\in\{0,1\}$ and each $C \in \exor \pr{t_{i-1},x_i,t_i}$ (if $i=1$, omit  $t_{i-1}$), the clause $A_{i,C}^b\in \psi_i$ yields the line $(C\cup\{a_{i,n}\}, M_{n+1}^{1-b})$. Resolving each of these with each of $B_{i,n}^{d}$ for $d\in \{0,1\}$, we obtain four clauses that can be resolved in two pairs to produce the lines $(C\cup\{a_{i,n-1}\},M_n^b)$.  (See the derivation at the end of this proof.) Repeating this process successively for $j = n, n-1, \ldots, i+2$, using the clause pairs $B_{i,j}^d$ with the previously derived clauses, we can obtain each $(C\cup\{a_{i,j}\},M_{j+1}^b)$. In each stage, the index $j$ of the variable $a_{i,j}$ present in the clause decreases, while the merge map accounts for one more variable. Finally, when we use the clause pairs $B_{i,i+1}^d$, the $a_{i,i+1}$ variable is eliminated, variables $x_{i+1}, \ldots , x_n$ are accounted for in the merge map,  and we obtain the lines $(C,M_{i+1}^b)$, corresponding to the clauses in $\widetilde{\psi_i}$. 
  
The derivation at one stage is as shown below.   
	{\small \begin{prooftree}
		\AxiomC{$\pr{C \cup \br{a_{i,j}}, M^1_{j+1}}$}
		\AxiomC{$\overbrace{\pr{\br{\overline{a_{i,j}}, x_j, a_{i,j-1}},*}}^{B_{i,j}^0}$}
		\BinaryInfC{$\pr{C \cup \br{x_j, a_{i,j-1}},M^1_{j+1}}$}
		\AxiomC{$\pr{C \cup \br{a_{i,j}}, M^0_{j+1}}$}
		\AxiomC{$\overbrace{\pr{\br{\overline{a_{i,j}}, \overline{x_j}, a_{i,j-1}},*}}^{B_{i,j}^1}$}
		\BinaryInfC{$\pr{C \cup \br{\overline{x_j}, a_{i,j-1}},M^0_{j+1}}$}
		\BinaryInfC{$\pr{C \cup \br{a_{i,j-1}},M^1_{j}}$}
	\end{prooftree}}
	{\small \begin{prooftree}
		\AxiomC{$\pr{C \cup \br{a_{i,j}}, M^1_{j+1}}$}
		\AxiomC{$\overbrace{\pr{\br{\overline{a_{i,j}}, \overline{x_j}, a_{i,j-1}},*}}^{B_{i,j}^1}$}
		\BinaryInfC{$\pr{C \cup \br{\overline{x_j}, a_{i,j-1}},M^1_{j+1}}$}
		\AxiomC{$\pr{C \cup \br{a_{i,j}}, M^0_{j+1}}$}
		\AxiomC{$\overbrace{\pr{\br{\overline{a_{i,j}}, x_j, a_{i,j-1}},*}}^{B_{i,j}^0}$}
		\BinaryInfC{$\pr{C \cup \br{x_j, a_{i,j-1}},M^0_{j+1}}$}
		\BinaryInfC{$\pr{C \cup \br{a_{i,j-1}},M^0_{j}}$}
	\end{prooftree}}
\end{proof}

In the second phase, we successively eliminate the $t$ variables in stages. 
\begin{lem} \label{lem:mparity-derive-smaller-t}
	Let the quantifier prefix be as in the definition of $\MParity$.
	Then the following derivations can be done in $\MRes$ in size polynomial in $n$:
	\begin{enumerate}
		\item For $i = n, n-1, \ldots, 2$, $\pr{\br{t_i},M^1_{i+1}}, \pr{\br{\overline{t_i}},M^0_{i+1}}, \widetilde{\psi_{i}} \vdash \pr{\br{t_{i-1}},M^1_{i}}, \pr{\br{\overline{t_{i-1}}}, M^0_{i}}$.
		\item $\pr{\br{t_1},M^1_{2}}, \pr{\br{\overline{t_1}},M^0_{2}}, \widetilde{\psi_{1}} \vdash \pr{\square,M^1_1}$.
	\end{enumerate}
\end{lem}
\begin{proof}
For $i\ge 2$, the derivation is as follows:
	{\small \begin{prooftree}
		\AxiomC{$\pr{\br{t_{i-1}, x_i, \overline{t_i}},M^1_{i+1}}$}
		\AxiomC{$\pr{\br{t_i},M^1_{i+1}}$}
		\BinaryInfC{$\pr{\br{t_{i-1}, x_i}, M^1_{i+1}}$}
		\AxiomC{$\pr{\br{t_{i-1}, \overline{x_i}, t_i},M^0_{i+1}}$}
		\AxiomC{$\pr{\br{\overline{t_i}},M^0_{i+1}}$}
		\BinaryInfC{$\pr{\br{t_{i-1}, \overline{x_i}}, M^0_{i+1}}$}
		\BinaryInfC{$\pr{\br{t_{i-1}},M^1_{i}}$}
	\end{prooftree}}
	{\small \begin{prooftree}
		\AxiomC{$\pr{\br{\overline{t_{i-1}}, \overline{x_i}, \overline{t_i}},M^1_{i+1}}$}
		\AxiomC{$\pr{\br{t_i},M^1_{i+1}}$}
		\BinaryInfC{$\pr{\br{\overline{t_{i-1}}, \overline{x_i}},M^1_{i+1}}$}
		\AxiomC{$\pr{\br{\overline{t_{i-1}}, x_i, t_i},M^0_{i+1}}$}
		\AxiomC{$\pr{\br{\overline{t_i}},M^0_{i+1}}$}
		\BinaryInfC{$\pr{\br{\overline{t_{i-1}}, x_i},M^0_{i+1}}$}
		\BinaryInfC{$\pr{\br{\overline{t_{i-1}}},M^0_{i}}$}
	\end{prooftree}}
	
	The  derivation at the last stage is as follows:
	{\small \begin{prooftree}
		\AxiomC{$\pr{\br{x_1, \overline{t_1}},M^1_{2}}$}
		\AxiomC{$\pr{\br{t_1},M^1_{2}}$}
		\BinaryInfC{$\pr{\br{x_1}, M^1_{2}}$}
		\AxiomC{$\pr{\br{\overline{x_1}, t_1},M^0_{2}}$}
		\AxiomC{$\pr{\br{\overline{t_1}},M^0_{2}}$}
		\BinaryInfC{$\pr{\br{\overline{x_1}}, M^0_{2}}$}
		\BinaryInfC{$\pr{\square,M^1_{1}}$}
	\end{prooftree}}
\end{proof}

We can now conclude the following:
\begin{lem}\label{lem:mparity-mres-poly}
	$\MParity$ has polynomial-size $\MRes$ refutations.
\end{lem}
\begin{proof}
  We first use \autoref{lem:mparity-add-complex-strategies} to derive all the $\widetilde{\psi_i}$. 
Next, we  start with $\pr{\br{t_n},M^1_{n+1}}$ and $ \pr{\br{\overline{t_n}},M^0_{n+1}}$, the lines corresponding to the clauses in $\psi_{n+1}$. From these lines  and $\widetilde{\psi_{n}}$, we derive $\pr{\br{t_{n-1}},M^1_{n}}$ and $ \pr{\br{\overline{t_{n-1}}},M^0_{n}}$, using \autoref{lem:mparity-derive-smaller-t}. We  continue in this manner deriving $\pr{\br{t_i},M^1_{i+1}}$ and $\pr{\br{\overline{t_i}},M^0_{i+1}}$ for $i = n-2, n-3, \ldots, 1$. From the lines $\widetilde{\psi_{1}}$, $\pr{\br{t_1},M^1_{2}}$ and $\pr{\br{\overline{t_1}},M^0_{2}}$, we  derive $\pr{\square,M^1_{1}}$ using \autoref{lem:mparity-derive-smaller-t}. 
\end{proof}

\begin{thm}\label{thm:adv-over-lquplus}
	 $\LDQRes$ does not p-simulate $\MRes$; and
	 $\LQURes$ and $\LQUplusRes$ are incomparable with $\MRes$.
\end{thm}

\begin{proof}
	We showed in \autoref{lem:mparity-mres-poly} that the $\MParity$ formulas have polynomial-size $\MRes$ refutations.
	We will now show that $\MParity$ requires exponential-size $\LQUplusRes$ refutations. We first note that $\QUParity$ requires exponential-size $\LQUplusRes$ refutations \cite{BCJ-ToCT-Sep2019}. We further note that $\LQUplusRes$ is closed under restrictions (Proposition 2 in \cite{BWJ-SAT14}). Since restricting the $\MParity$ formulas by setting $a_{i,j}=0$, for all $i,j \in[n]$, gives the $\QUParity$ formulas, we conclude that $\MParity$ requires exponential-size $\LQUplusRes$ refutations. Therefore $\LQUplusRes$ does not simulate $\MRes$.
	Since $\LQUplusRes$ p-simulates $\LDQRes$ and $\LQURes$, these two systems also do not simulate $\MRes$.

        In \cite{BBMPS-ToCT}  it is shown that $\MRes$ does not simulate $\QURes$. (The separating formula is in fact $\KBKFlq$.) Since $\LQURes$ and $\LQUplusRes$ p-simulate $\QURes$ \cite{BWJ-SAT14} and the simulation order is transitive, it follows that $\MRes$ does not simulate $\LQURes$ and $\LQUplusRes$. 

Hence $\LQURes$ and $\LQUplusRes$ are incomparable with $\MRes$.
\end{proof}

\begin{rem}
	In these proofs, note that the hardness for $\LQUplusRes$ and $\IRM$ was proven using restrictions. But the same did not apply to $\MRes$ --- a restricted formula being hard for $\MRes$ does not mean that the original formula is also hard. This means that $\MRes$ is not closed under restrictions, and is hence unnatural.
\end{rem}

\begin{rem}
	Another observation is that the clauses of the $\KBKFlqweak$ formula family are weakenings of the clauses of $\KBKFlq$. Since $\KBKFlq$ requires exponential-size $\MRes$ refutations but $\KBKFlqweak$ has polynomial-size $\MRes$ refutations, we conclude that weakening adds power to $\MRes$.
\end{rem}

\section{Role of weakenings, and unnaturalness}
\subsection{Weakenings} \label{subsec:weakenings}
Let $(C, \{M^u \mid u \in U\})$ be a line of an $\MRes$ refutation. Then it can be weakened in two different ways \cite{BBM-JAR20}:
\begin{itemize}
	\item{Existential clause weakening:} $C \vee x$ can be derived from $C$, provided it does not contain the literal $\overline{x}$. The merge maps remain the same. Similarly, $C\vee \overline{x}$ can be derived if $x \not\in C$. 
	\item {Strategy weakening:} A trivial merge map ($*$) can be replaced by a constant merge map ($0$ or $1$). The existential clause remains the same.
\end{itemize}

Adding these weakenings to $\MRes$ gives the following three proof systems:
\begin{itemize}
	\item $\MRes$ with existential clause weakening ($\MResWE$),
	\item $\MRes$ with strategy weakening ($\MResWU$), and
	\item $\MRes$ with both existential clause and strategy weakening ($\MResWEU$).
\end{itemize}
\begin{propC}[\cite{BBM-JAR20}]
All the proof systems $\MResWE$, $\MResWU$, and $\MResWEU$ are complete and sound. 
\end{propC}
\begin{proof}
  By definition, $\MResWEU$ $p$-simulates $\MResWE$ and $\MResWU$,
  both of which $p$-simulate $\MRes$.  Completeness thus follows from
  the completeness of $\MRes$. It suffices to show that $\MResWEU$ is
  sound. This follows from Lemma~31 in \cite{BBM-JAR20}, where a
  generalized version of this system, allowing simultaneous
  existential clause weakening and universal strategy weakening in a
  single rule application, is shown to be sound for the Herbrand-form
  Dependency QBFs (H-form DQBFs, discussed in Section~6 of
  \cite{BBM-JAR20}). Since QBFs are a special case of H-form DQBFs, soundness of $\MResWEU$ for QBFs follows. 
\end{proof}

In the remainder of this subsection, we will study the relation among these systems.

First, we note that existential clause weakening adds exponential power. 
\begin{thm} \label{thm:mreswe-vs-mres}
	$\MResWE$ is strictly stronger than $\MRes$.
\end{thm}
\begin{proof}
	Since $\MResWE$ is a generalization of $\MRes$, $\MResWE$ p-simulates $\MRes$.
	
	The $\KBKFlq$ formulas can be transformed  into the $\KBKFlqweak$ formulas in $\MResWE$ using a linear number of applications of the existential weakening rule. The transformed $\KBKFlqweak$ formulas have polynomial-size $\MRes$ (and hence $\MResWE$) refutations, \autoref{lem:kbkfweak-mres-poly}. Thus the $\KBKFlq$ formulas have polynomial-size $\MResWE$ refutations. Since the $\KBKFlq$ formulas require exponential-size $\MRes$ refutations \cite{BBMPS-ToCT}, we get the desired separation.
\end{proof}

Next we observe that a lower bound for $\MRes$,  Theorem 3.17  from \cite{BBMPS-ToCT}, can be lifted to $\MResWU$. 
\begin{restatable}{restate-lemma}{restateMResWULowerBound}\label{lem:kbkflq-mreswu-hard}
	$\KBKFlq$ requires exponential-size refutations in $\MResWU$.
\end{restatable}
\begin{proof}
	We observe that the	$\MRes$ lower bound for $\KBKFlq$ in \cite{BBMPS-ToCT} works with a minor modification. In \cite[Lemma 3.19]{BBMPS-ToCT}, item 3 says that $M^{x_i} = *$. However a weaker condition $M^{x_i} \in \{*,0,1\}$ is sufficient for the lower bound.
	With this modification, we observe that the remaining argument carries over, and the lower bound also works for $\MResWU$.
	
	For the convenience of the reader, we reproduce this proof from \cite{BBMPS-ToCT} in \autoref{sec:kbkflq-mreswu-hard-complete-proof} with the modification incorporated.
\end{proof}
\begin{thm} \label{thm:strategy-vs-exist-weak}
	$\MResWU$ does not simulate $\MResWE$; and
	$\MResWEU$ is strictly stronger than $\MResWU$.
\end{thm}

\begin{proof}%
	We showed that the $\KBKFlq$ formulas require refutations of exponential size in $\MResWU$ (\autoref{lem:kbkflq-mreswu-hard}) but have polynomial-size refutations in $\MResWE$ and $\MResWEU$ (proof of \autoref{thm:mreswe-vs-mres}). Therefore $\MResWU$ does not simulate $\MResWE$ and $\MResWEU$. Since $\MResWEU$ p-simulates $\MResWU$, $\MResWEU$ is strictly stronger than $\MResWU$.
\end{proof}

The next logical question is whether strategy weakening adds power to $\MRes$. We do not know the answer. However, we can answer this for the regular versions of these systems.

\begin{defi}
	A refutation is called regular if each variable is resolved at most once along every leaf-to-root path. A proof system is called regular if it only allows regular refutations.
\end{defi}

\begin{thm} \label{thm:strategy-weak-adds-power-to-reg-mres}
	Regular $\MResWU$ is strictly stronger than regular $\MRes$.
\end{thm}
To prove this theorem, 
we will use a variant of the Squared-Equality ($\SquaredEquality$) formula family, called Squared-Equality-with-Holes ($\SquaredEqualityWithHoles$). Squared-Equality, defined in \cite{BBM-JAR20}, is a two-dimensional version of the Equality ($\Equality$) formula family \cite{BBH-LMCS19}, and has short regular tree-like $\MRes$ refutations. It was used to show that the systems $\QRes$, $\QURes$, reductionless $\LDQRes$, $\ExpRes$, $\IR$ and $\QCP$ do not p-simulate $\MRes$. We recall the definitions of Equality and Squared-Equality formulas below:

\begin{defi}
	\emph{Equality} ($\Equality(n)$) is the following QBF formula family:
	\[
	\displaystyle\mathop{\mathlarger{\exists}}_{i \in [n]} x_i,  \mathop{\mathlarger{\forall}}_{i \in [n]} u_i,  \mathop{\mathlarger{\exists}}_{i \in [n]} t_{i}.\; \pr{\mathop{\mathlarger{\wedge}}_{i \in [n]} A_{i}} \wedge B
	\]
	where
	\begin{itemize}
	\item $B = \displaystyle \vee_{i \in [n]} \overline{t_{i}}$,
	\item For $i \in [n]$, $A_{i}$ contains the following two clauses:
	\begin{align*}
		x_i \vee u_i \vee t_{i}, & &\overline{x_i} \vee \overline{u_i} \vee t_{i}.
	\end{align*}
\end{itemize}
\end{defi}

To see that the Equality formulas are false, 
observe that $u_i=x_i$ for each $i\in[n]$ is a winning strategy for the universal player. 
If the universal player plays according to this strategy, then  after $x$ and $u$ are set, for  each $i \in [n]$, one of the clauses in $A_i$ becomes $t_i$. These reduced clauses, along with $B$, cannot all be satisfied by the existential plater.

\begin{defi}\label{def:squared-equality}
	\emph{Squared-Equality} ($\SquaredEquality(n)$) is the following QBF family:
	\[
	\displaystyle\mathop{\mathlarger{\exists}}_{i \in [n]} x_i, \mathop{\mathlarger{\exists}}_{j \in [n]} y_j, \mathop{\mathlarger{\forall}}_{i \in [n]} u_i, \mathop{\mathlarger{\forall}}_{j \in [n]} v_j, \mathop{\mathlarger{\exists}}_{i,j \in [n]} t_{i,j}.\; \pr{\mathop{\mathlarger{\wedge}}_{i,j \in [n]} A_{i,j}} \wedge B
	\]
	where
	\begin{itemize}
		\item $B = \displaystyle \vee_{i,j \in [n]} \overline{t_{i,j}}$,
		\item For $i,j \in [n]$, $A_{i,j}$ contains the following four clauses:
		\begin{align*}
& x_i \bigvee y_j \bigvee u_i \bigvee v_j \bigvee t_{i,j}, &
& x_i \bigvee \overline{y_j} \bigvee u_i \bigvee \overline{v_j} \bigvee t_{i,j},\\
& \overline{x_i} \bigvee y_j \bigvee \overline{u_i} \bigvee v_j \bigvee t_{i,j}, &
& \overline{x_i} \bigvee \overline{y_j} \bigvee \overline{u_i} \bigvee \overline{v_j} \bigvee t_{i,j}
		\end{align*}
	\end{itemize}
\end{defi}

For the Squared Equality formulas, the following strategy for the universal player is a winning strategy: for all $i \in [n]$, set $u_i = x_i$; and for all $j \in [n]$, set $v_j = y_j$.

We observe that the short $\MRes$ refutation of $\SquaredEquality(n)$ crucially uses the isomorphism of merge maps. For each $i,j \in [n]$, the four clauses in $A_{i,j}$ are resolved to derive the line $\pr{t_{i,j},\br{u_i=x_i,v_j=y_j}}$. These lines are then resolved with the line $\pr{\vee_{i,j \in [n]} \overline{t_{i,j}}, \br{*, \cdots, *}}$ to derive the line $\pr{\square, \br{u_i=x_i, v_j=y_j \mid \forall i,j \in [n]}}$. The resolutions over the $t_{i,j}$ variables are possible only because the merge maps are isomorphic. If we modify the clauses of $\SquaredEquality$ such that the merge maps produced from different $A_{i,j}$ are non-isomorphic, then the refutation described above is forbidden. This is the motivation behind the Squared-Equality-with-Holes ($\SquaredEqualityWithHoles$) formula family defined below. It is constructed from $\SquaredEquality$ by removing some of the universal variables from the $A_{i,j}$ clauses. The resulting QBF family remains false but different $A_{i,j}$ lead to different merge maps. We believe that this QBF family is hard for $\MRes$, but we have only been able to prove the hardness for regular $\MRes$, and hence the separation is between the regular versions of $\MRes$ and $\MResWU$.

The variant identifies regions in the $[n]\times[n]$ grid, and changes the clause sets $A_{i,j}$ depending on the region that $(i,j)$ belongs to. We can use any partition of $[n]\times [n]$ into two regions $R_0,R_1$ such that each region has at least one position in each row and at least one position in each column; call such a partition a \emph{covering partition}. One possible choice for $R_0$ and $R_1$ is the following: $R_0 = ([1,n/2] \times [1,n/2]) \cup ([n/2+1,n] \times [n/2+1,n])$ and $R_1 = ([1,n/2] \times [n/2+1,n]) \cup ([n/2+1,n] \times [1,n/2])$. We will call $R_0$ and $R_1$ the two regions of the matrix.

\begin{defi}\label{def:h-sq-eq}
  Let $R_0,R_1$ be a covering partition of $[n]\times [n]$. 

  Squared-Equality-with-Holes ($\SquaredEqualityWithHoles(n)(R_0,R_1)$) is the following QBF family:
	\[
	\displaystyle\mathop{\mathlarger{\exists}}_{i \in [n]} x_i, y_i, \mathop{\mathlarger{\forall}}_{j \in [n]} u_j, v_j, \mathop{\mathlarger{\exists}}_{i,j \in [n]} t_{i,j}.\; \pr{\mathop{\mathlarger{\wedge}}_{i,j \in [n]} A_{i,j}} \wedge B
	\]
	where
	
	\begin{itemize}
		\item $B = \displaystyle \vee_{i,j \in [n]} \overline{t_{i,j}}$,
		\item For $(i,j) \in R_0$, $A_{i,j}$ contains the following four clauses:
		\begin{align*}
			& x_i \vee y_j \vee u_i \vee v_j \vee t_{i,j},  &
			& x_i \vee \overline{y_j} \vee u_i \vee t_{i,j},\\
			& \overline{x_i} \vee y_j \vee v_j \vee t_{i,j},  &
			& \overline{x_i} \vee \overline{y_j} \vee t_{i,j}
		\end{align*}
		\item For $(i,j) \in R_1$, $A_{i,j}$ contains the following four clauses:
		\begin{align*}
			& x_i \vee y_j \vee t_{i,j}, &
			&  x_i \vee \overline{y_j} \vee \overline{v_j} \vee t_{i,j},\\
			& \overline{x_i} \vee y_j \vee \overline{u_i} \vee t_{i,j}, &
			&  \overline{x_i} \vee \overline{y_j} \vee \overline{u_i} \vee \overline{v_j} \vee t_{i,j}
		\end{align*}
	\end{itemize}
(We do not always specify the regions explicitly but merely say  $\SquaredEqualityWithHoles$.)
\end{defi}

\begin{lem} \label{lem:h-sq-eq-lb}
	$\SquaredEqualityWithHoles(n)$  requires exponential-size refutations in regular $\MRes$.
\end{lem}
Before proving this, we show how to obtain  \autoref{thm:strategy-weak-adds-power-to-reg-mres}.
\begin{proof}[Proof of \autoref{thm:strategy-weak-adds-power-to-reg-mres}]
	Since regular $\MResWU$ is a generalization of regular $\MRes$, it p-simulates regular $\MRes$.
	
	Using strategy weakening, we can get $\SquaredEquality$ from $\SquaredEqualityWithHoles$ in a linear number of steps. Since $\SquaredEquality$ has polynomial-size refutations in regular $\MRes$, we get polynomial-size refutations for $\SquaredEqualityWithHoles$ in regular $\MResWU$.	On the other hand, \autoref{lem:h-sq-eq-lb} gives an exponential lower bound for $\SquaredEqualityWithHoles$ in regular $\MRes$. Therefore regular $\MResWU$ is strictly stronger than regular $\MRes$.
\end{proof}

It remains to prove \autoref{lem:h-sq-eq-lb}.
This is a fairly involved proof, but in broad outline and in many details it is similar to the lower bound for $\SquaredEquality$ in reductionless $\LDQRes$ 
\cite{BBM-JAR20}. 

The size bound is trivially true for $n=1$, so we assume that $n > 1$. Let $\Pi$ be a Regular $\MRes$ refutation of $\SquaredEqualityWithHoles(n)$. Since a tautological clause cannot occur in a regular $\MRes$ refutation, we assume that $\Pi$ does not have a line whose clause part is tautological.

Let us first fix some notation. Let $X = \{x_1, \ldots, x_n\}$, $Y = \{y_1, \ldots, y_n\}$, $U = \{u_1, \ldots, u_n\}$, $V = \{v_1, \ldots, v_n\}$, and $T = \{t_{i,j} \mid i,j \in [n]\}$.
For lines $L_1$, $L_2$, etc., the respective clauses and merge maps will be denoted by $C_1$, $C_2$ and $M_1$, $M_2$ etc.
For a line $L$ in $\Pi$, $\Pi_L$ denotes the sub-derivation of $\Pi$ ending in $L$. Viewing  $\Pi$ as a directed acyclic graph, we can talk of leaves and paths in $\Pi$. 
For a line $L$ of $\Pi$, let $\UsedConstraintInd(L) = \br{ (i,j) \mid A_{i,j} \cap \leaves(\Pi_L) \neq \emptyset }$. (The abbreviation $\UsedConstraintInd$ stands for UsedConstraintsIndex.)

We first show some structural properties about $\Pi$. The first property excludes using many axioms in certain derivations. 
\begin{lem} \label{lem:UCI}
  For line  $L = (C, M)$  of $\Pi$, and $i,j\in [n]$, if $t_{i,j} \in C$, then $\UsedConstraintInd(L) = \br{(i,j)}$.
  
\end{lem}
\begin{proof}
	Since the literal $t_{i,j}$ only occurs in clauses in $A_{i,j}$, so $\leaves(L) \cap A_{i,j} \neq \emptyset$, hence $\UsedConstraintInd(L) \supseteq \br{(i,j)}$.

	Now suppose $\card{\UsedConstraintInd(L)} > 1$. Let $(i',j')$ be an arbitrary element of $\UsedConstraintInd(L)$ distinct from  $(i,j)$.
Pick a leaf of $\Pi_L$ using a clause in $A_{i',j'}$, and let $p$ be a path from this leaf to $L$ and then to the final line of $\Pi$. Both $t_{i,j}$ and $t_{i',j'}$ are necessarily used as pivots on this path. Assume that $t_{i,j}$ is used as a pivot later (closer to the final line) than $t_{i',j'}$; the other case is symmetric. Let $L_c = \res(L_a, L_b, t_{i',j'})$ and $L_f = \res(L_d, L_e, t_{i,j})$ respectively be the positions where $t_{i',j'}$ and $t_{i,j}$ are used as resolution pivots on this path (here $L_a$ and $L_d$ are the lines of path $p$, hence $t_{i',j'} \in C_a$ and $t_{i,j} \in C_d$). Then $C_b$ has the negated literal $\overline{t_{i',j'}}$; hence $B \in \leaves(L_b)$. Since $\overline{t_{i,j}} \in B$ but $\overline{t_{i,j}} \notin L_d$, $t_{i,j}$ is used as a resolution pivot in the derivation $\Pi_{L_d}$. This contradicts the fact that $\Pi$ is regular.
\end{proof}

The next property is the heart of the proof, and shows that paths with $B$ at the leaf must have a suitable wide clause. 
\begin{lem} \label{lem:h-eq-wide-clause}
	On every path from $\pr{\vee_{i,j \in [n]} \overline{t_{i,j}}, \br{*, \cdots, *}}$ (the line for axiom clause $B$) to the final line, there exists a line $L = (C, M)$ such that either $X \subseteq var(C)$ or $Y \subseteq var(C)$.
\end{lem}
\begin{proof}
  With each line $L_l = (C_l, M_l)$ in $\Pi$, we associate an $n \times n$ matrix $N_l$ in which $N_l[i,j] = 1$ if $\overline{t_{i,j}} \in C_l$ and $N_l[i,j] = 0$ otherwise.
  
  Let $p = L_1, \ldots, L_k$ be a path from $\pr{\vee_{i,j \in [n]} \overline{t_{i,j}}, \br{*, \cdots, *}}$ to the final line in $\Pi$. Since $\Pi$ is regular, each $\overline{t_{i,j}}$ is resolved away exactly once, so no clause on $p$ has any positive $t_{i,j}$ literal.
  Let $l$ be the least integer such that $N_l$ has a $0$ in each row or a $0$ in each column. Note that $l \geq 2$ since $N_1$ has no zeros.	Consider the case  that $N_l$ has a $0$ in each row; the argument for the other case is identical. We will show in this case that $X \subseteq \var(C_l)$. We will use the following claim:
	
	\begin{clm} \label{claim:EachRowZeroAndOne}
		In each row of $N_l$, there is a $0$ and a $1$ such that the $0$ and $1$ are in different regions (i.e.~one is in $R_0$ and the other in $R_1$).
	\end{clm}	
\begin{proof}
	We already know that $N_l$ has a $0$ in each row. We will first prove that $N_l$ also has a $1$ in each row. Aiming for contradiction, suppose that $N_l$ has a full $0$ row $r$. Since $l \geq 2$, $N_{l-1}$ exists. Note that, by definition of resolution, there can be at most one element that changes from $1$ in $N_{l-1}$ to $0$ in $N_l$. Since $N_{l-1}$ does not have a $0$ in every column, it does not contain a full $0$ row. Hence, the unique element that changed from $1$ in $N_{l-1}$ to $0$ in $N_l$ must be in row $r$. Thus all other rows of $N_{l-1}$ already contain the one $0$ of that row in $N_l$. Since $n \geq 2$, $N_{l-1}$ also has at least one $0$ in row $r$; thus $N_{l-1}$ has a $0$ in each row, contradicting the minimality of $l$.

Since $R_0$ and $R_1$ form a covering partition, it cannot be the case that all the $0$s and $1$s of any row are in the same region $R_b$; that would imply that $R_{1-b}$ does not cover the row.
(Proof of \autoref{claim:EachRowZeroAndOne} complete.)\end{proof}

We want to prove that $X \subseteq var(C_l)$. Suppose, to the contrary,  there exists $i \in [n]$ such that $x_i \notin \var(C_l)$. We know that
	there exist $j_1, j_2 \in [n]$ such that $N_l[i,j_1] = 0$ and $N_l[i,j_2] = 1$; and either $(i,j_1) \in R_0$ and $(i,j_2) \in R_1$, or $(i,j_1) \in R_1$ and $(i,j_2) \in R_0$. Without loss of generality, we may assume that $(i,j_1) \in R_0$ and $(i,j_2) \in R_1$.
	
	We know that on path $p$, there is a resolution with pivot $t_{i,j_1}$ before $L_l$ and a resolution with pivot $t_{i,j_2}$ after $L_l$. Let the former resolution be $L_c = \res(L_a, L_b, t_{i,j_1})$ where $L_b$ is on path $p$, and let the latter resolution be $L_f = \res(L_d, L_e, t_{i,j_2})$ where $L_e$ is on path $p$. Since $\Pi$ is a regular refutation, $t_{i,j_1} \in C_a$, $\overline{t_{i,j_1}} \in C_b$ and $t_{i,j_2} \in C_d$, $\overline{t_{i,j_2}} \in C_e$. Thus along path $p$ these lines appear in the relative order $B, L_b, L_c, L_l, L_e, L_f, (\Box, \{u_i = x_i, v_j = y_j \mid \forall i,j \in [n]\} )$. 

        \begin{clm}\label{claim:negated-x}
          $\overline{x_i} \in C_c$.
        \end{clm}
        \begin{proof}
	By \autoref{lem:UCI}, $\UsedConstraintInd(L_d) = \{(i,j_2)\}$, or equivalently $\leaves(L_d) \subseteq A_{i,j_2}$. Since $(i,j_2) \in R_1$, no clause in $A_{i,j_2}$ has literal $u_i$. Hence $M_d^{u_{i}} \in \{*,1\}$. Furthermore, if $M_d^{u_{i}} = *$, then $x_i \in C_d$. Since the pivot for resolving $L_d$ and $L_e$ is $t_{i,j_2}$, this would imply that $x_i \in C_f$.

	By a similar argument, we can conclude that
	\begin{enumerate*}[label=(\roman*)]
		\item $\leaves(L_a) \subseteq A_{i,j_1}$,
		\item $M_a^{u_{i}} \in \{*,0\}$, and
		\item if $M_a^{u_{i}} = *$, then $\overline{x_i} \in C_c$.
	\end{enumerate*}

        If $M_d^{u_i}=*$ and $M_a^{u_i}=*$, then $x_i\in C_f$  and $\overline{x_i}\in C_c$. So $x_i$ must be used twice as a pivot on path $p$, contradicting regularity. 
        
        If $M_d^{u_i}=*$ and $M_a^{u_i}=0$, then $x_i\in C_f$ and $\Pi_{L_a}$ uses some clause containing $x_i$ to make the merge map for $u_i$ non-trivial. Thus $x_i \in \Pi_{L_a}$, $x_i\not\in L_l$ by assumption, $x_i\in L_f$. Hence $x_i$ is used twice as pivot, contradicting regularity.

        Hence $M_d^{u_i}=1$. Since the resolution at line $L_f$ is not blocked, $M_e^{u_i} \in\{*,1\}$. But $L_e$ is derived after, and using, $L_a$. Since merge maps do not get simpler along a path,  $M_a^{u_i} \in\{*,1\}$. It follows that  $M_a^{u_i} = *$.
 Hence $\overline{x_i} \in C_c$. 
(Proof of \autoref{claim:negated-x} complete.)	\end{proof}
        
	Since $\overline{x_i} \notin C_l$, $x_i$ has been used as a resolution pivot between $L_c$ and $L_l$ on path $p$. Let $L_w = \res(L_u, L_v, x_i)$ be the position on path $p$ where $x_i$ is used as pivot (since the refutation is regular, such a position is unique). Let $L_v$ be the line on path $p$. By regularity of the refutation, $x_i \in L_u$ and $\overline{x_i} \in L_v$.

As observed at the outset, $L_w$ is on path $p$ and so does not contain a positive $t$ literal. Since $C_w$ is obtained via pivot $x_i$, this implies that $C_u$ also does not contain a positive $t$ literal. Since all axioms contain at least one $t$ variable but only $B$ contains negated $t$ literals, so $B \in \leaves(L_u)$.
	
Let $q$ be a path that starts from a leaf using $B$, passes through $L_u$ to $L_w$, and then continues along path $p$ to the final clause.
Since the refutation is regular, $N_v = N_u = N_w$. Hence $N_v[i,j_1] = 0$ i.e.~$\overline{t_{i,j_1}} \notin C_v$. This implies that $t_{i,j_1}$ is used as resolution pivot before $L_v$ on path $q$.

We already know that $t_{i,j_2}$ is used as a pivot after line $L_l$ on path $p$, and hence on path $q$.
Arguing analogous to \autoref{claim:negated-x}  for path $p$ but with respect to path $q$, we observe that $\overline{x_i}$ belongs to at least one leaf of $L_u$. Since $x_i \in C_u$ and since the refutation is regular, $x_i$ is not used as a resolution pivot before $C_u$ on path $q$. This implies that $\overline{x_u} \in C_u$. We already know that $x_i \in C_u$, since it contributed the pivot at $L_w$. This means that $C_u$ is a tautological clause, a contradiction. 
\end{proof}

We can finally prove \autoref{lem:h-sq-eq-lb}. This part is identical to the corresponding part of the proof of Theorem 28 in \cite{BBM-JAR20}; we include it here for completeness. 
\begin{proof}[Proof of \autoref{lem:h-sq-eq-lb}]
	For each $a = \pr{a_1, \ldots, a_n} \in \br{0,1}^n$, consider the assignment $\sigma_a$ to the existential variables which sets $x_i = y_i = a_i$ for all $i \in [n]$, and $t_{i,j} = 1$ for all $i,j \in [n]$. Call such an assignment a symmetric assignment. Given a symmetric assignment $\sigma_a$, walk from the final line of $\Pi$ towards the leaves maintaining the following invariant: for each line $L = (C, \{M^u \mid u \in U \cup V\})$, $\sigma_a$ falsifies $C$. Let $p_a$ be the path followed. By \autoref{lem:h-eq-wide-clause}, this path will contain a line $L = (C, \{M^u \mid u \in U \cup V\})$ such that either $X \subseteq var(C)$ or $Y \subseteq var(C)$. Let us define a function $f$ from symmetric assignments to the lines of $\Pi$ as follows: $f(a) = (C, \{M^u \mid u \in U \cup V\})$ is the last line (i.e.~nearest to the leaves) on $p_a$ such that either $X \subseteq var(C)$ or $Y \subseteq var(C)$. Note that, for any line $L$ of $\Pi$, there can be at most one symmetric assignment $a$ such that $f(a) = L$. This means that there are at least $2^{n}$ lines in $\Pi$. This gives the desired lower bound.
\end{proof}

\subsection{Simulation by \texorpdfstring{$\QBFeFrege$}{eFrege with universal reduction}}
It was recently shown that $\QBFeFrege$ p-simulates all known resolution-based QBF proof systems; in particular, it p-simulates $\MRes$ \cite{ChewSlivovsky-LMCS24}. We observe that this p-simulation can be extended in a straightforward manner to handle both the weakenings in $\MRes$. Hence we obtain a p-simulation of $\MResWE$, $\MResWU$ and $\MResWEU$ by $\QBFeFrege$.

\begin{thm} \label{thm:QBFeFrege-simulates-MResW}
	$\QBFeFrege$ strictly p-simulates $\MResWE$, $\MResWU$ and $\MResWEU$.
\end{thm}
\begin{proof}%
The separation follows from the separation of the propositional proof systems resolution and $\eFrege$ \cite{Urquhart-JACM87}. We prove the p-simulation below.

It suffices to prove that $\QBFeFrege$ p-simulates $\MResWEU$. %
The proof is essentially same as that of the p-simulation of $\MRes$ in \cite{ChewSlivovsky-LMCS24}, but with two additional cases for the two weakenings. So, we will briefly describe that proof and then describe the required modifications.

Let $\Pi$ be an $\MResWEU$ refutation $\Pi$ of a QBF $\Phi$. The last line of this refutation gives a winning strategy for the universal player; let us call this strategy $S$. We will first prove that there is a short $\eFrege$ derivation $\Phi \vdash \neg S$. 	
	Then, as mentioned in \cite{ChewSlivovsky-LMCS24}, the technique of \cite{BeyersdorffBCP-JACM20,Chew-SAT21} can be used to derive the empty clause from $\neg S$ using universal reduction.
	
	We will now describe an $\eFrege$ derivation $\Phi \vdash \neg S$. Let $L_i = \pr{C_i, \br{M^u_i \mid u \in U}}$ be the i\textsuperscript{th} line of $\Pi$. We create new extension variables: $s^u_{i,j}$ is the variable for the j\textsuperscript{th} node of $M^u_i$. If node $j$ is a leaf of $M^u_i$ labelled by constant $c$, then $s^u_{i,j}$ is defined to be $c$. Otherwise, if $M^u_i (j) = (x,a,b)$, then $s^u_{i,j}$ is defined as $s^u_{i,j} \triangleq \big(x \wedge s^u_{i,a}\big) \vee \big(\overline{x} \wedge s^u_{i,b}\big)$. The extension variables for $u$ will be to its left in the quantifier prefix.
	
	We will prove that for each line $L_i$ of $\Pi$, we can derive the formula $F_i \triangleq \wedge_{u \in U_i} (u \leftrightarrow s^u_{i, r(u,i)}) \rightarrow C_i$; where $r(u,i)$ is the index of the root of merge map $M^u_i$, and $U_i$ is the set of universal variables for which $M^u_i$ is non-trivial.
	
	Our proof will proceed by induction on the lines of the refutation.
	
	The base case is when $L_i$ is an axiom; and the inductive step will have three cases depending on which rule is used to derive $L_i$:
	\begin{enumerate*}[label=(\roman*)]
		\item resolution,
		\item existential clause weakening, or
		\item strategy weakening.
	\end{enumerate*}
	The proof for the base case and the resolution step case is  as given in \cite{ChewSlivovsky-LMCS24}. We give proofs for the other two cases below:
	\begin{itemize}
	\item{Existential clause weakening:} Let line $L_b = \pr{C_b, \br{M^u_b \mid u \in U}}$ be derived from line $L_a = \pr{C_a, \br{M^u_a \mid u \in U}}$ using existential clause weakening. Then $C_b = C_a \vee x$ for some existential literal $x$ such that $\overline{x} \notin C_a$, and $M^u_b = M^u_a$ for all $u \in U$. By the induction hypothesis, we have derived the formula $F_a\triangleq \wedge_{u \in U_a} (u \leftrightarrow s^u_{a, r(u,a)}) \rightarrow C_a$. We have to derive the formula $F_b\triangleq\wedge_{u \in U_b} (u \leftrightarrow s^u_{b, r(u,b)}) \rightarrow C_b=\wedge_{u \in U_b} (u \leftrightarrow s^u_{b, r(u,b)}) \rightarrow C_a \vee x$. Since $M^u_b = M^u_a$ for each $u$, there is a short $\QBFeFrege$ derivation of the formula $s^u_{a,j} \leftrightarrow s^u_{b,j}$ for each $u \in U_i$, and each node $j$ of $M_a^u$. This allows us to replace variable $s^u_{a,j}$ by $s^u_{b,j}$ in $F_a$. As a result, we get the formula $F'_b\triangleq \wedge_{u \in U_b} (u \leftrightarrow s^u_{b, r(u,b)}) \rightarrow C_a$. Now, using an inference of the form $p \rightarrow q \vdash p \rightarrow q \vee r$, we obtain the formula $F_b$.

	\item{Strategy weakening:} 
          Let line $L_b = \pr{C_b, \br{M^u_b \mid u \in U}}$ be derived from line $L_a = \pr{C_a, \br{M^u_a \mid u \in U}}$ using strategy weakening for a variable $v$. Then $C_b = C_a$, $M^u_b = M^u_a$ for all $u \in U\setminus\{v\}$, and $M_a^v=*,$ $M_b^v$ is a constant, say $d$.
Similar to the above case, we start with the inductively obtained $F_a$ and replace each $s_{a,j}^u$ with $s_{b,j}^u$ to obtain a formula $F'_b\triangleq \wedge_{u \in U_b\setminus\{v\}} (u \leftrightarrow s^u_{b, r(u,b)}) \rightarrow C_b$. With a final  inference  of the form $p \rightarrow q \vdash p \wedge r \rightarrow q$, we can then add $(v \leftrightarrow s^v_{b,r(v,b)})$ to the conjunction to obtain $F_b$. \qedhere
	\end{itemize}
\end{proof}

\subsection{Unnaturalness}
In this section, we observe  that $\MRes$ and $\MResWU$ are unnatural proof systems, i.e.\ they are not closed under restrictions.

\begin{thm}\label{thm:unnatural}
	$\MRes$ and $\MResWU$ are unnatural proof systems.
\end{thm}
\begin{proof}
  The $\KBKFlqsplit$ formula family has polynomial-size refutations in $\MRes$ (and $\MResWU$), as seen in \autoref{lem:kbkfsplit-mres-poly}. The restriction of this family obtained by setting $t=0$ is exactly the $\KBKFlq$ formula family, which, as shown in \autoref{lem:kbkflq-mreswu-hard}, is exponentially hard for $\MResWU$ and hence also for $\MRes$.
\end{proof}

\section{Conclusion and future work}
$\MRes$ was introduced in \cite{BBM-JAR20} to overcome the weakness of $\LDQRes$. It was shown that $\MRes$ has advantages over many proof systems, but the advantage over $\LDQRes$ was not demonstrated. In this paper, we have filled this gap. We have shown that $\MRes$ has advantages over not only $\LDQRes$, but also over more powerful systems, $\LQUplusRes$ and $\IRM$. While we also know that $\MRes$ cannot simulate $\LQUplusRes$, it remains open whether $\MRes$ can actually simulate $\LDQRes$ or even $\IRM$. We have also looked at the role of weakening and shown that it adds power to $\MRes$. 

On the negative side, we have shown that $\MRes$ with and without strategy weakening is unnatural --- which could well make it useless in practice. 
It is possible, but yet unproven, that $\MRes$  can be made natural by adding existential weakening or both weakenings. This, in our opinion, is the most important open question about $\MRes$. However, it is worth noting that even for propositional SAT solving, the most successful solvers are based on determinizations of CDCL, which is not natural (closed under restrictions), and so unnaturalness may not be a real hindrance after all.

\section*{Acknowledgment}
We thank Olaf Beyersdorff, Joshua Blinkhorn, and Tom\'{a}\v{s} Peitl for interesting discussions about the power of $\MRes$. Part of this work was done at Schloss Dagstuhl Leibniz Centre for Informatics during seminar 20061 (SAT and Interactions).

\bibliographystyle{alphaurl}%
\bibliography{mergeres}

\newcommand{\etalchar}[1]{$^{#1}$}
\begin{thebibliography}{MMZ{\etalchar{+}}01}

\bibitem[Ats04]{Atserias-JACM04}
Albert Atserias.
\newblock On sufficient conditions for unsatisfiability of random formulas.
\newblock {\em J. {ACM}}, 51(2):281--311, 2004.
\newblock \href {https://doi.org/10.1145/972639.972645} {\path{doi:10.1145/972639.972645}}.

\bibitem[BBCP20]{BeyersdorffBCP-JACM20}
Olaf Beyersdorff, Ilario Bonacina, Leroy Chew, and J{\'{a}}n Pich.
\newblock Frege systems for quantified {Boolean} logic.
\newblock {\em J. {ACM}}, 67(2):9:1--9:36, 2020.
\newblock \href {https://doi.org/10.1145/3381881} {\path{doi:10.1145/3381881}}.

\bibitem[BBH19]{BBH-LMCS19}
Olaf Beyersdorff, Joshua Blinkhorn, and Luke Hinde.
\newblock Size, cost, and capacity: {A} semantic technique for hard random {QBFs}.
\newblock {\em Log. Methods Comput. Sci.}, 15(1), 2019.
\newblock \href {https://doi.org/10.23638/LMCS-15(1:13)2019} {\path{doi:10.23638/LMCS-15(1:13)2019}}.

\bibitem[BBM21]{BBM-JAR20}
Olaf Beyersdorff, Joshua Blinkhorn, and Meena Mahajan.
\newblock Building strategies into {QBF} proofs.
\newblock {\em J. Autom. Reason.}, 65(1):125--154, 2021.
\newblock {Preliminary version in the proceedings of the STACS 2019, LIPIcs vol.\ 126, 14:1--14:18}.
\newblock \href {https://doi.org/10.1007/s10817-020-09560-1} {\path{doi:10.1007/s10817-020-09560-1}}.

\bibitem[BBM{\etalchar{+}}24]{BBMPS-ToCT}
Olaf Beyersdorff, Joshua Blinkhorn, Meena Mahajan, Tom\'{a}\v{s} Peitl, and Gaurav Sood.
\newblock {Hard QBFs for Merge Resolution}.
\newblock {\em ACM Trans. Comput. Theory}, 16(2), March 2024.
\newblock {Preliminary version in the proceedings of FSTTCS 2020, LIPIcs vol.\ 182, 12:1--12:15}.
\newblock \href {https://doi.org/10.1145/3638263} {\path{doi:10.1145/3638263}}.

\bibitem[BCJ19]{BCJ-ToCT-Sep2019}
Olaf Beyersdorff, Leroy Chew, and Mikol{\'{a}}s Janota.
\newblock New resolution-based {QBF} calculi and their proof complexity.
\newblock {\em {ACM} Trans. Comput. Theory}, 11(4):26:1--26:42, 2019.
\newblock \href {https://doi.org/10.1145/3352155} {\path{doi:10.1145/3352155}}.

\bibitem[BCMS18]{BCMS-IC18}
Olaf Beyersdorff, Leroy Chew, Meena Mahajan, and Anil Shukla.
\newblock Understanding cutting planes for {QBFs}.
\newblock {\em Inf. Comput.}, 262:141--161, 2018.
\newblock \href {https://doi.org/10.1016/j.ic.2018.08.002} {\path{doi:10.1016/j.ic.2018.08.002}}.

\bibitem[BJ12]{BJ-FMSD12}
Valeriy Balabanov and Jie{-}Hong~R. Jiang.
\newblock Unified {QBF} certification and its applications.
\newblock {\em Formal Methods Syst. Des.}, 41(1):45--65, 2012.
\newblock \href {https://doi.org/10.1007/s10703-012-0152-6} {\path{doi:10.1007/s10703-012-0152-6}}.

\bibitem[BKS04]{BKS-JAIR04}
Paul Beame, Henry~A. Kautz, and Ashish Sabharwal.
\newblock Towards understanding and harnessing the potential of clause learning.
\newblock {\em J. Artif. Intell. Res.}, 22:319--351, 2004.
\newblock \href {https://doi.org/10.1613/jair.1410} {\path{doi:10.1613/jair.1410}}.

\bibitem[Bla37]{Blake-Thesis37}
Archie Blake.
\newblock {\em Canonical expressions in {B}oolean algebra}.
\newblock PhD thesis, University of Chicago, 1937.

\bibitem[BN21]{BussN-HandbookofSAT21}
Sam Buss and Jakob Nordstr{\"o}m.
\newblock Proof complexity and {SAT} solving.
\newblock In Armin Biere, Marijn Heule, Hans van Maaren, and Toby Walsh, editors, {\em Handbook of Satisfiability}, pages 233--350. IOS Press, Netherlands, 2nd edition, May 2021.
\newblock \href {https://doi.org/10.3233/FAIA200990} {\path{doi:10.3233/FAIA200990}}.

\bibitem[BPS21]{BlinkhornPS-SAT21}
Joshua Blinkhorn, Tom{\'{a}}s Peitl, and Friedrich Slivovsky.
\newblock {Davis} and {Putnam} meet {Henkin}: Solving {DQBF} with resolution.
\newblock In Chu{-}Min Li and Felip Many{\`{a}}, editors, {\em Theory and Applications of Satisfiability Testing - {SAT} 2021 - 24th International Conference, Barcelona, Spain, July 5-9, 2021, Proceedings}, volume 12831 of {\em Lecture Notes in Computer Science}, pages 30--46. Springer, 2021.
\newblock \href {https://doi.org/10.1007/978-3-030-80223-3\_4} {\path{doi:10.1007/978-3-030-80223-3\_4}}.

\bibitem[BWJ14]{BWJ-SAT14}
Valeriy Balabanov, Magdalena Widl, and Jie{-}Hong~R. Jiang.
\newblock {QBF} resolution systems and their proof complexities.
\newblock In Carsten Sinz and Uwe Egly, editors, {\em Theory and Applications of Satisfiability Testing - {SAT} 2014 - 17th International Conference, Held as Part of the Vienna Summer of Logic, {VSL} 2014, Vienna, Austria, July 14-17, 2014. Proceedings}, volume 8561 of {\em Lecture Notes in Computer Science}, pages 154--169. Springer, 2014.
\newblock \href {https://doi.org/10.1007/978-3-319-09284-3\_12} {\path{doi:10.1007/978-3-319-09284-3\_12}}.

\bibitem[Che17]{LeroyChew-Thesis}
Leroy~Nicholas Chew.
\newblock {\em {QBF} proof complexity}.
\newblock PhD thesis, University of Leeds, 2017.
\newblock URL: \url{https://etheses.whiterose.ac.uk/18281/}.

\bibitem[Che21]{Chew-SAT21}
Leroy Chew.
\newblock Hardness and optimality in {QBF} proof systems modulo {NP}.
\newblock In Chu{-}Min Li and Felip Many{\`{a}}, editors, {\em Theory and Applications of Satisfiability Testing - {SAT} 2021 - 24th International Conference, Barcelona, Spain, July 5-9, 2021, Proceedings}, volume 12831 of {\em Lecture Notes in Computer Science}, pages 98--115. Springer, 2021.
\newblock \href {https://doi.org/10.1007/978-3-030-80223-3\_8} {\path{doi:10.1007/978-3-030-80223-3\_8}}.

\bibitem[CR79]{CR-JSL79}
Stephen~A. Cook and Robert~A. Reckhow.
\newblock The relative efficiency of propositional proof systems.
\newblock {\em Journal of Symbolic Logic}, 44(1):36--50, 1979.

\bibitem[CS23]{CS-STACS23}
Sravanthi Chede and Anil Shukla.
\newblock Extending merge resolution to a family of {QBF}-proof systems.
\newblock In Petra Berenbrink, Patricia Bouyer, Anuj Dawar, and Mamadou~Moustapha Kant{\'{e}}, editors, {\em 40th International Symposium on Theoretical Aspects of Computer Science, {STACS} 2023, March 7-9, 2023, Hamburg, Germany}, volume 254 of {\em LIPIcs}, pages 21:1--21:20. Schloss Dagstuhl - Leibniz-Zentrum f{\"{u}}r Informatik, 2023.
\newblock URL: \url{https://doi.org/10.4230/LIPIcs.STACS.2023.21}, \href {https://doi.org/10.4230/LIPICS.STACS.2023.21} {\path{doi:10.4230/LIPICS.STACS.2023.21}}.

\bibitem[CS24]{ChewSlivovsky-LMCS24}
Leroy Chew and Friedrich Slivovsky.
\newblock Towards uniform certification in {QBF}.
\newblock {\em Log. Methods Comput. Sci.}, 20(1), 2024.
\newblock {Preliminary version in the proceedings of STACS 2022, LIPIcs vol.\ 219, 22:1--22:23}.
\newblock \href {https://doi.org/10.46298/LMCS-20(1:14)2024} {\path{doi:10.46298/LMCS-20(1:14)2024}}.

\bibitem[DLL62]{DavisLL-CACM62}
Martin Davis, George Logemann, and Donald~W. Loveland.
\newblock A machine program for theorem-proving.
\newblock {\em Commun. {ACM}}, 5(7):394--397, 1962.
\newblock \href {https://doi.org/10.1145/368273.368557} {\path{doi:10.1145/368273.368557}}.

\bibitem[DP60]{DavisP-JACM60}
Martin Davis and Hilary Putnam.
\newblock A computing procedure for quantification theory.
\newblock {\em J. {ACM}}, 7(3):201--215, 1960.
\newblock \href {https://doi.org/10.1145/321033.321034} {\path{doi:10.1145/321033.321034}}.

\bibitem[ELW13]{EglyLW-LPAR13}
Uwe Egly, Florian Lonsing, and Magdalena Widl.
\newblock Long-distance resolution: Proof generation and strategy extraction in search-based {QBF} solving.
\newblock In Kenneth~L. McMillan, Aart Middeldorp, and Andrei Voronkov, editors, {\em Logic for Programming, Artificial Intelligence, and Reasoning - 19th International Conference, LPAR-19, Stellenbosch, South Africa, December 14-19, 2013. Proceedings}, volume 8312 of {\em Lecture Notes in Computer Science}, pages 291--308. Springer, 2013.
\newblock \href {https://doi.org/10.1007/978-3-642-45221-5\_21} {\path{doi:10.1007/978-3-642-45221-5\_21}}.

\bibitem[Gel12]{Gelder-CP12}
Allen~Van Gelder.
\newblock Contributions to the theory of practical quantified boolean formula solving.
\newblock In Michela Milano, editor, {\em Principles and Practice of Constraint Programming - 18th International Conference, {CP} 2012, Qu{\'{e}}bec City, QC, Canada, October 8-12, 2012. Proceedings}, volume 7514 of {\em Lecture Notes in Computer Science}, pages 647--663. Springer, 2012.
\newblock \href {https://doi.org/10.1007/978-3-642-33558-7\_47} {\path{doi:10.1007/978-3-642-33558-7\_47}}.

\bibitem[HK17]{HeuleK-CACM17}
Marijn J.~H. Heule and Oliver Kullmann.
\newblock The science of brute force.
\newblock {\em Commun. {ACM}}, 60(8):70--79, 2017.
\newblock \href {https://doi.org/10.1145/3107239} {\path{doi:10.1145/3107239}}.

\bibitem[JM15]{Janota-Expansion-vs-QRes-TCS15}
Mikol{\'{a}}s Janota and Jo{\~{a}}o Marques{-}Silva.
\newblock Expansion-based {QBF} solving versus {Q}-resolution.
\newblock {\em Theor. Comput. Sci.}, 577:25--42, 2015.
\newblock \href {https://doi.org/10.1016/j.tcs.2015.01.048} {\path{doi:10.1016/j.tcs.2015.01.048}}.

\bibitem[KKF95]{KKF-IC95}
Hans Kleine{ }B{\"{u}}ning, Marek Karpinski, and Andreas Fl{\"{o}}gel.
\newblock Resolution for quantified boolean formulas.
\newblock {\em Inf. Comput.}, 117(1):12--18, 1995.
\newblock \href {https://doi.org/10.1006/inco.1995.1025} {\path{doi:10.1006/inco.1995.1025}}.

\bibitem[Kra19]{krajicek2019}
Jan Krajíček.
\newblock {\em Proof Complexity}.
\newblock Encyclopedia of Mathematics and its Applications. Cambridge University Press, 2019.
\newblock \href {https://doi.org/10.1017/9781108242066} {\path{doi:10.1017/9781108242066}}.

\bibitem[MMZ{\etalchar{+}}01]{MMZZM-CDCL-DAC01}
Matthew~W. Moskewicz, Conor~F. Madigan, Ying Zhao, Lintao Zhang, and Sharad Malik.
\newblock Chaff: Engineering an efficient {SAT} solver.
\newblock In {\em Proceedings of the 38th Design Automation Conference, {DAC} 2001, Las Vegas, NV, USA, June 18-22, 2001}, pages 530--535. {ACM}, 2001.
\newblock \href {https://doi.org/10.1145/378239.379017} {\path{doi:10.1145/378239.379017}}.

\bibitem[MS22]{MS-SAT22}
Meena Mahajan and Gaurav Sood.
\newblock {QBF} merge resolution is powerful but unnatural.
\newblock In Kuldeep~S. Meel and Ofer Strichman, editors, {\em 25th International Conference on Theory and Applications of Satisfiability Testing, {SAT} 2022, August 2-5, 2022, Haifa, Israel}, volume 236 of {\em LIPIcs}, pages 22:1--22:19. Schloss Dagstuhl - Leibniz-Zentrum f{\"{u}}r Informatik, 2022.
\newblock \href {https://doi.org/10.4230/LIPICS.SAT.2022.22} {\path{doi:10.4230/LIPICS.SAT.2022.22}}.

\bibitem[MSLM21]{MLM-HandbookofSAT21}
Joao Marques-Silva, Ines Lynce, and Sharad Malik.
\newblock Conﬂict-driven clause learning {SAT} solvers.
\newblock In Armin Biere, Marijn Heule, Hans van Maaren, and Toby Walsh, editors, {\em Handbook of Satisfiability}, pages 133--182. IOS Press, Netherlands, 2nd edition, May 2021.
\newblock \href {https://doi.org/10.3233/FAIA200987} {\path{doi:10.3233/FAIA200987}}.

\bibitem[Rob65]{Robinson-JACM60}
John~Alan Robinson.
\newblock A machine-oriented logic based on the resolution principle.
\newblock {\em J. {ACM}}, 12(1):23--41, 1965.
\newblock \href {https://doi.org/10.1145/321250.321253} {\path{doi:10.1145/321250.321253}}.

\bibitem[SBPS19]{ShuklaBPS-ICTAI19}
Ankit Shukla, Armin Biere, Luca Pulina, and Martina Seidl.
\newblock A survey on applications of quantified {Boolean} formulas.
\newblock In {\em 31st {IEEE} International Conference on Tools with Artificial Intelligence, {ICTAI} 2019, Portland, OR, USA, November 4-6, 2019}, pages 78--84. {IEEE}, 2019.
\newblock \href {https://doi.org/10.1109/ICTAI.2019.00020} {\path{doi:10.1109/ICTAI.2019.00020}}.

\bibitem[SM73]{StockmeyerM-STOC73}
Larry~J. Stockmeyer and Albert~R. Meyer.
\newblock Word problems requiring exponential time: Preliminary report.
\newblock In Alfred~V. Aho, Allan Borodin, Robert~L. Constable, Robert~W. Floyd, Michael~A. Harrison, Richard~M. Karp, and H.~Raymond Strong, editors, {\em Proceedings of the 5th Annual {ACM} Symposium on Theory of Computing, April 30 - May 2, 1973, Austin, Texas, {USA}}, pages 1--9. {ACM}, 1973.
\newblock \href {https://doi.org/10.1145/800125.804029} {\path{doi:10.1145/800125.804029}}.

\bibitem[SS99]{MSS-CDCL-ITC99}
Jo{\~{a}}o P.~Marques Silva and Karem~A. Sakallah.
\newblock {GRASP:} {A} search algorithm for propositional satisfiability.
\newblock {\em {IEEE} Trans. Computers}, 48(5):506--521, 1999.
\newblock \href {https://doi.org/10.1109/12.769433} {\path{doi:10.1109/12.769433}}.

\bibitem[Urq87]{Urquhart-JACM87}
Alasdair Urquhart.
\newblock Hard examples for resolution.
\newblock {\em J. {ACM}}, 34(1):209--219, 1987.
\newblock \href {https://doi.org/10.1145/7531.8928} {\path{doi:10.1145/7531.8928}}.

\bibitem[Var14]{Vardi-Satisfiability-CACM14}
Moshe~Y. Vardi.
\newblock Boolean satisfiability: theory and engineering.
\newblock {\em Commun. {ACM}}, 57(3):5, 2014.
\newblock \href {https://doi.org/10.1145/2578043} {\path{doi:10.1145/2578043}}.

\bibitem[ZM02]{ZhangM-ICCAD02}
Lintao Zhang and Sharad Malik.
\newblock Conflict driven learning in a quantified {Boolean} satisfiability solver.
\newblock In Lawrence~T. Pileggi and Andreas Kuehlmann, editors, {\em Proceedings of the 2002 {IEEE/ACM} International Conference on Computer-aided Design, {ICCAD} 2002, San Jose, California, USA, November 10-14, 2002}, pages 442--449. {ACM} / {IEEE} Computer Society, 2002.
\newblock \href {https://doi.org/10.1145/774572.774637} {\path{doi:10.1145/774572.774637}}.

\end{thebibliography}

\appendix

\section{Complete proof of \autoref{lem:kbkflq-mreswu-hard}} \label{sec:kbkflq-mreswu-hard-complete-proof}
As mentioned in \autoref{subsec:weakenings}, the proof of the $\MRes$ lower bound for $\KBKFlq$ formulas \cite[Theorem 3.17]{BBMPS-ToCT} can be generalized to $\MResWU$ by doing a minor modification to the proof. In this section, we sketch the proof of \cite[Theorem 3.17]{BBMPS-ToCT}, highlighting the appropriate modification needed to handle strategy weakening. Most of the text is reproduced directly from \cite{BBMPS-ToCT}.

Recall the definition of the $\KBKFlq$ formulas from \autoref{def:KBKFlq}.
Note that the existential part of each clause in $\KBKFlq_n$ is a
Horn clause (at most one positive literal), and except $A_0$, is even
strict Horn (exactly one positive literal).

We use the following shorthand notation.
Sets of variables: $D =  \{d_1, \ldots, d_n\}$, $E =  \{e_1, \ldots, e_n\}$, $F =  \{f_1, \ldots, f_n\}$, and 
$X =  \{x_1, \ldots, x_n\}$.
Sets of literals: For $Y\in \{D,E,X,F\}$, set 
$Y^1 = \{ u \mid u\in Y\}$ and
$Y^0 = \{ \overline{u} \mid u\in Y\}$. 
Sets of clauses:
\[\begin{array}[t]{lclclcl}
	\mathcal{A}_0 &=& \{A_0\}\\
	\mathcal{A}_i &=& \{A^d_i, A^e_i\} \quad \forall i \in [n] & \quad \quad & \mathcal{B}_i &=& \{B^0_i, B^1_i\} \quad \forall i \in [n]\\
	\mathcal{A}_{[i,j]} &=& \cup_{k \in [i,j]} \mathcal{A}_k \quad \forall i,j \in [0,n], i \le j & \quad & \mathcal{B}_{[i,j]} &=& \cup_{k \in [i,j]} \mathcal{B}_k \quad \forall i,j \in [n], i \le j\\
	\mathcal{A} &=& \mathcal{A}_{[0,n]} & \quad & \mathcal{B} &=& \mathcal{B}_{[1,n]} 
\end{array}
\]

\autoref{lem:kbkflq-mreswu-hard} asserts that $\KBKFlq$ requires exponential-size to refute in $\MResWU$. 
The proof follows the following high-level idea. Let $\Pi$ be a $\MResWU$
refutation of $\KBKFlq$. Since every axiom of $\KBKFlq$ contains a variable from $F$ while the final clause of $\Pi$ is empty, there is a maximal ``component'' (say $\mathcal{S}$) of $\Pi$ leading to and including the final line, where all clauses are $F$-free. The clauses in this component only contain variables in $D$ and $E$. We show that the ``boundary'' ($\partial{\mathcal{S}}$) of this component is large. To show that the boundary is large, \cite{BBMPS-ToCT} identify a property of merge maps called {\em self-dependence} which captures the right complexity; a merge map for $x_i$ has this self-dependence property if it depends on at least one of $d_i,e_i$. We show that all merge maps at the final line must have self-dependence, whereas at the  boundary lines none of the merge maps have self-dependence. 
We use this to then  conclude that there must be exponentially many lines.

To show that self-dependence is not possible outside the $F$-free
component, we show that from a line with $F$-variables and at least
one self-dependent strategy, the $F$-variables can never be removed.

Elaborating on the roadmap of the argument: 
Let $\Pi$ be an $\MResWU$ refutation of $\KBKFlq_n$.  Each line in $\Pi$
has the form $L=(C,M^{x_1},\ldots,M^{x_n})$ where $C$ is a clause over
$D,E,F$, and each $M^{x_i}$ is a merge map computing a strategy for $x_i$.

Define $\mathcal{S}$ to be the set of those lines in $\Pi$ where the
clause part has no $F$ variable and furthermore the line has a path in
$G_\Pi$ to the final empty clause via lines where all the clauses also
have no $F$ variables. Let $\partial{\mathcal{S}}$, called the boundary of $\mathcal{S}$, denote the set of leaves in
the subgraph of $G_\Pi$ restricted to $\mathcal{S}$; these are lines
that are in $\mathcal{S}$ but their parents are not in
$\mathcal{S}$.  Note that by definition, for each $L=(C,\{M^{x_i}\mid
i\in[n]\}) \in \mathcal{S}$, $\var(C) \subseteq D \cup E$. No line in
$\mathcal{S}$ (and in particular, no line in $\partial{\mathcal{S}}$) is an
axiom since all axiom clauses have variables from $F$.

Recall that the variables of $\KBKFlq_n$ can be naturally grouped
based on the quantifier prefix: for $i\in[n]$, the $i$th group has
$d_i,e_i, x_i$, and the $(n+1)$th group has the $F$ variables. By
construction, the merge map for $x_i$ does not depend on variables in
later groups, as is indeed required for a countermodel.  We say that a
merge map for $x_i$ has {\em self-dependence} if it does depend on $d_i$
and/or $e_i$.

We show that every merge map at every line in $\mathcal{S}$ is non-trivial
(\autoref{lem:S'-strategies-nontrivial}).  Further, we show that at
every line on the boundary of $\mathcal{S}$, i.e.~in $\partial{\mathcal{S}}$, no merge map has
self-dependence (\autoref{lem:boundary-strategies}). Using this, we
conclude that $\partial{\mathcal{S}}$ must be exponentially large, since in every
countermodel the strategy of each variable must have self-dependence
(\autoref{prop:KBKF-strategies}).

In order to show that lines in $\partial{\mathcal{S}}$ do not have self-dependence, we first
establish several properties of the sets of axiom clauses used in a
sub-derivation
(Lemmas \ref{lem:posFliterals-and-uci}, \ref{lem:empty-uci-structure}, \ref{lem:nonempty-uci-structure} and \ref{lem:self-dependent-strategy-suffix-interval}).

For a line $L \in \Pi$, let $\Pi_{L}$ be the minimal sub-derivation of
$L$, and let $G_{\Pi_{L}}$ be the corresponding subgraph of $G_{\Pi}$
with sink $L$. Let $\UsedConstraintInd(\Pi_{L}) = \{ i \in [0,n] \mid
\leaves(G_{\Pi_{L}}) \cap \mathcal{A}_i \neq
\emptyset\}$. ($\UsedConstraintInd$ stands for
UsedConstraintsIndex). Note that we are only looking at the clauses in
$\mathcal{A}$ to define $\UsedConstraintInd$.

\begin{lemC}[\texorpdfstring{\cite[Lemma 3.18]{BBMPS-ToCT}}{Lemma 3.18}]
	\label{lem:posFliterals-and-uci}
	For every line $L = (C, \{M^{x_i} \mid i \in [n]\})$ of $\Pi$, 
	$\card{C\cap F^1}\le 1$. Furthermore,
	$\UsedConstraintInd(\Pi_L) = \emptyset\Leftrightarrow C \cap F^1 \neq \emptyset$. 
	(Here, $F^1 = \{f_1, f_2, \ldots, f_n\}$ is the set of positive literals over the variable set $\{f_1, f_2, \ldots, f_n\}$.)
\end{lemC}

\begin{lem}[adapted from \texorpdfstring{\cite[Lemma 3.19]{BBMPS-ToCT}}{Lemma 3.19}]
	\label{lem:empty-uci-structure}
	A line $L = (C, \{M^{x_i} \mid i \in [n]\})$ of $\Pi$ with
	$\UsedConstraintInd(\Pi_L) = \emptyset$ has these properties:
	\begin{enumerate}
		\item \label{item:empty-uci-vars}
		$\var(C) \subseteq F$; for all  $i\in[n]$, $M^{x_i} \in \{*,0,1\}$;
		\item \label{item:empty-uci-positive-map}
		For some $j\in [n]$, $f_j \in C$ and $M^{x_j} \in \{0,1\}$; such a $j$ is unique; 
		\item \label{item:empty-uci-maps-left-of-positive}
		For the unique $j$ from (\ref{item:empty-uci-positive-map}), for $1 \le i < j$, $f_i \not\in \var(C)$ and $M^{x_i} \in \{*,0,1\}$;
		\item \label{item:empty-uci-maps-right-of-positive}
		For $j < i \le n$,
		if $f_i\not\in \var(C)$, then $M^{x_j} \in \{0,1\}$.
	\end{enumerate}
\end{lem}
{\bf This statement differs from \cite[Lemma 3.19]{BBMPS-ToCT} in exactly one respect:} in \autoref{item:empty-uci-maps-left-of-positive} we have a weaker conclusion  (already shown in \autoref{item:empty-uci-vars}) that $M^{x_i} \in \{*,0,1\}$, 
whereas in \cite[Lemma 3.19]{BBMPS-ToCT} it was further  proved that $M^{x_i}=*$. Since strategy weakening allows us to replace a * by 0 or 1, we cannot draw this conclusion.  However, the stronger conclusion was not used to prove the subsequent \autoref {item:empty-uci-maps-right-of-positive}, so nothing changes in the proof.

In subsequent lemmas, whenever we use this lemma, we need to show that the weaker conclusion suffices. It turns out that subsequent lemmas use this item essentially  to say that the merge map is not complex; it is trivial $*$ or simple $0,1$. This conclusion remains valid with the modified version.

\begin{lemC}[\texorpdfstring{\cite[Lemma 3.20]{BBMPS-ToCT}}{Lemma 3.20}]
	\label{lem:nonempty-uci-structure}  
	Let $L = (C, \{M^{x_i} \mid i \in [n]\})$ be a line of $\Pi$ with
	$\UsedConstraintInd(\Pi_L) \neq \emptyset$.  Then
	$\UsedConstraintInd(\Pi_{L})$ is an interval $[a,b]$ for some $0 \le
	a \le b \le n$. Furthermore, {\upshape(}in the items below, $a,b$ refer to the
	endpoints of this interval\,{\upshape)}, it 
	has the following properties:  
	\begin{enumerate}
		\item \label{item:no-*-in-uci}
		For $k\in[n]\cap [a,b]$, $M^{x_k}\neq *$.
		\item \label{item:start-uci-positive-literals}
		If $a \ge 1$, then $\card{\{d_a,e_a\}\cap C}=1$.
		If $a=0$, then $C$ does not have any positive literal.
		\item \label{item:end-uci-negated-literals}
		If $b < n$, then $\overline{d_{b+1}}, \overline{e_{b+1}} \in C$.
		\item \label{item:outside-uci-mergemaps}
		For all $k \in [n]\setminus[a,b]$, {\upshape(i)}~$d_k, e_k \not\in \var(M^{x_{k}})$, and 
		{\upshape(ii)}~if $M^{x_k} = *$ then $\overline{f_k} \in C$.
	\end{enumerate}
\end{lemC}

\begin{lemC}[\texorpdfstring{\cite[Lemma 3.21]{BBMPS-ToCT}}{Lemma 3.21}]
	\label{lem:self-dependent-strategy-suffix-interval}
	For any line $L=(C,\{M^{x_i}\mid i\in[n]\})$ in $\Pi$, and any
	$k\in[n]$, if $\{d_k,e_k\} \cap \var(M^{x_k}) \neq \emptyset$, then 
	$\UsedConstraintInd(\Pi_{L}) =[a,n]$ for some $a\le k-1$.
\end{lemC}

\begin{lemC}[\texorpdfstring{\cite[Lemma 3.22]{BBMPS-ToCT}}{Lemma 3.22}]
	\label{lem:S'-strategies-nontrivial}
	For all $L \in \mathcal{S}$, for all $k \in [n]$, $M^{x_{k}}
	\neq *$. 
\end{lemC}

\begin{lemC}[\texorpdfstring{\cite[Lemma 3.23]{BBMPS-ToCT}}{Lemma 3.23}]
	\label{lem:boundary-strategies}
	For all $L \in \partial{\mathcal{S}}$, for all $k \in [n]$,
	$d_k, e_k \not\in \var(M^{x_{k}})$. 
\end{lemC}

We will also use the following property of $\KBKFlq$ formulas. It
implies that in every countermodel, the strategy for every variable
has self-dependence. This is used, towards the end of the proof of
\autoref{lem:kbkflq-mreswu-hard}, to show that merge maps for countermodels must be complex and large. 

\begin{propC}[\texorpdfstring{\cite[Proposition 3.24]{BBMPS-ToCT}}{Proposition 3.24}]
	\label{prop:KBKF-strategies}
	Let $h$ be any countermodel for $\KBKFlq_n$.  Let $\alpha$ be any
	assignment to $D$, and $\beta$ be any assignment to $E$. For
	each $i\in [n]$, if $\alpha_j \neq \beta_j$ for all $1\le j
	\le i$, then $h^{x_i}\big((\alpha,\beta)\restriction_{L_Q(x_i)}\big) = \alpha_i$. In particular, if $\alpha_j\ne\beta_j$ for all $j\in[n]$,
	then the countermodel computes $h(\alpha,\beta)=\alpha$.
\end{propC}

Now we have all the required information; we put it together to obtain
the lower bound. 

\begin{proof}[Proof of \autoref{lem:kbkflq-mreswu-hard}]
	Let $\Pi$ be a refutation of $\KBKFlq_n$ in $\MResWU$. 
	Let $\mathcal{S},\partial{\mathcal{S}}$ be as defined in the beginning of
	this section.
	Let the final line of $\Pi$ be $L_\Box=(\Box,\{M_\Box^{x_i} \mid i\in
	[n]\})$, and for $i\in [n]$, let $h_i$ be the functions computed by
	the merge map $M_\Box^{x_i}$. By soundness of $\MResWU$, the functions
	$\{h_i\}_{i\in[n]}$ form a countermodel for $\KBKFlq_n$.
	
	For each $a \in \{0,1\}^n$, consider the assignment $\alpha$ to the
	variables of $D\cup E$ where $d_i=a_i$, $e_i = \overline{a_i}$. Call
	such an assignment an anti-symmetric assignment.  Given such an
	assignment, walk from $L_\Box$ towards the leaves of $\Pi$ as far as
	is possible while maintaining the following invariant at each line
	$L = (C,\{ M^{x_i}\mid i\in[n]\})$ along the way:
	\begin{enumerate}
		\item $\alpha$ falsifies $C$, and
		\item for each $i\in[n]$,  $h_i(\alpha) = M^{x_i}(\alpha)$.
	\end{enumerate}
	Clearly this invariant is initially true at $L_\Box$, which is
        in $\mathcal{S}$. If we are currently at a line $L \in
        \mathcal{S}$ where the invariant is true, and if $L\not\in
        \partial{\mathcal{S}}$, then consider how $L$ is derived. If
        it is obtained by using strategy weakening on some $L'$, then
        $C'=C$ and so $L'$ is also in $\mathcal{S}$. Then by
        \autoref{lem:S'-strategies-nontrivial}, each strategy in $L'$
        is already non-trivial, so no weakening is possible.  {\bf (Note:
        This argument, that there is no strategy weakening inside $S$,
        is the only addition needed to adapt the lower bound for
        $\MRes$ to $\MResWU$.) }

        Hence it must be the case that $L$ is derived using resolution. Say $L$ 
	is obtained from lines $L'$, $L''$. The resolution pivot in this step
	is not in $F$, since that would put $L$ in $\partial{\mathcal{S}}$. So both
	$L'$ and $L''$ are in $\mathcal{S}$, and the pivot is in $D \cup
	E$. Let the pivot be in $\{d_\ell,e_\ell\}$ for some
	$\ell\in[n]$. Depending on the pivot value, exactly one of $C',C''$
	is falsified by $\alpha$; say $C'$ is falsified.  By
	\autoref{lem:S'-strategies-nontrivial}, for each $i\in[n]$, both
	$(M')^{x_i}$ and $(M'')^{x_i}$ are non-trivial. By definition of the
	$\MRes$ rule,
	\begin{itemize}
		\item For $i< \ell$, $(M')^{x_i}$ and $(M'')^{x_i}$ are isomorphic
		(otherwise the resolution is blocked),
		and $M^{x_i} = (M')^{x_i} = (M'')^{x_i}$.
		\item For $i\ge \ell$, there are two possibilities: \\
		(1)~$(M')^{x_i}$ and $(M'')^{x_i}$ are isomorphic, and
		$M^{x_i} = (M')^{x_i}$.  \\
		(2)~$M^{x_i}$ is a merge of $(M')^{x_i}$
		and $(M'')^{x_i}$ with the pivot variable queried. By definition
		of the merge operation, since $C'$ is falsified by $\alpha$,
		$M^{x_i}(\alpha) = (M')^{x_i}(\alpha)$.
	\end{itemize}
	Thus in all cases, for each $i$, $h_i(\alpha) = M^{x_i}(\alpha)=
	(M')^{x_i}(\alpha)$.
	Hence  $L'$ satisfies the   invariant.
	
	We have shown that as long as we have not encountered a line in 
	$\partial{\mathcal{S}}$, we can move further. We continue the walk until a
	line in $\partial{\mathcal{S}}$ is reached. We denote the line so reached by
	$P(\alpha)$.  Thus $P$ defines a map from anti-symmetric assignments
	to $\partial{\mathcal{S}}$.

	We now show that the map $P$ is one-to-one. Suppose, to the
	contrary,
	$P(\alpha) = P(\beta) = (C,\{M^{x_i} \mid i\in[n]\})$ for
	two distinct anti-symmetric assignments obtained from $a,b\in
	\{0,1\}^n$ respectively. Let $j$ be the least index in $[n]$ where
	$a_j \neq b_j$.  By \autoref{lem:boundary-strategies}, $M^{x_j}$
	depends only on $\{d_i,e_i \mid i < j\}$, and $\alpha,\beta$ agree
	on these variables. Thus we get the equalities
	$a_j = h_j(\alpha) = M^{x_j}(\alpha) = M^{x_j}(\beta) = h_j(\beta)= b_j$,
	where the first and last equalities follow from
	\autoref{prop:KBKF-strategies}, the third equality from 
	\autoref{lem:boundary-strategies} and choice of $j$, and the second
	and fourth equalities by the invariant satisfied at $P(\alpha)$ and
	$P(\beta)$ respectively.
	This contradicts $a_j\neq b_j$.
	
	We have established that the map $P$ is one-to-one. Hence, 
	$\partial{\mathcal{S}}$ has at least as many lines as anti-symmetric
	assignments, so $\card{\Pi} \ge \card{\partial{\mathcal{S}}} \ge 2^n$.
\end{proof}

\end{document}